\documentclass[prl,floatfix,twocolumn,superscriptaddress]{revtex4-2}
\usepackage{amsmath}
\usepackage{amssymb}
\usepackage{amstext}
\usepackage{amsopn}
\usepackage{amsfonts}
\usepackage{amsxtra}
\usepackage[english]{babel}
\usepackage{graphicx}
\usepackage{float}
\usepackage{bm}
\usepackage{multirow}
\usepackage{dcolumn}
\usepackage{color}
\usepackage{hyperref}

\newcommand{\SIROx}{Sr$_2$Ir$_{1-x}$Rh$_{x}$O$_{4}$}
\newcommand{\SIO}{Sr$_2$IrO$_{4}$}
\newcommand{\LIO}{La$_3$Ir$_{3}$O$_{11}$}

\begin{document}

\title{\textbf{Orbital-Selective Spin-Orbit Mott Insulator in Fractional Valence Iridate La$_3$Ir$_3$O$_{11}$}}

\author{Kai Wang}
\email[]{kai.wang@henu.edu.cn}
\affiliation{School of Physics and Electronics, Henan University, Kaifeng 475004, China}
\affiliation{Institute of Quantum Materials and Physics, Henan Academy of Sciences, Zhengzhou 450046, China}

\author{Jun Yang}
\affiliation{Department of Materials Science and Engineering, Westlake University, Hangzhou, Zhejiang, China}

\author{Chaoyang Kang}
\affiliation{School of Future Technology, Henan University, Zhengzhou 450046, China}

\author{Weikang Wu}
\affiliation{Key Laboratory for Liquid-Solid Structural Evolution and Processing of Materials, Ministry of Education, Shandong University, Jinan 250061, China}

\author{Wenka Zhu}
\affiliation{High Magnetic Field Laboratory, Chinese Academy of Sciences, Hefei 230031, China}

\author{Jianzhou Zhao}
\email[]{jzzhao@tju.edu.cn}
\affiliation{Department of Physics, School of Science, Tianjin University, Tianjin 300354, China}

\author{Yaomin Dai}
\email[]{ymdai@nju.edu.cn}
\affiliation{National Laboratory of Solid State Microstructures and Department of Physics, Nanjing University, Nanjing 210093, China}

\author{Bing Xu}
\email[]{bingxu@iphy.ac.cn}
\affiliation{Beijing National Laboratory for Condensed Matter Physics, Institute of Physics, Chinese Academy of Sciences, P.O. Box 603, Beijing 100190, China}
\affiliation{School of Physical Sciences, University of Chinese Academy of Sciences, Beijing 100049, China}

\date{\today}
%
%

\begin{abstract}
The combination of strong spin-orbit coupling and Coulomb interactions makes the $5d$ iridates a unique platform for realizing novel correlated electronic states. Here, utilizing infrared spectroscopy, we demonstrate that a robust Mott insulating state persists in the $1/3$-hole self-doped system La$_3$Ir$_3$O$_{11}$, evidenced by the collapse of the Drude response and the emergence of sharp excitations across the Mott gap. Our theoretical calculations reveal that the insulating behavior arises from the cooperative interplay of structural distortions, spin-orbit coupling, and Coulomb interactions. Specifically, octahedral distortion and Ir-Ir dimerization split the $t_{2g}$ orbitals, driving the $J_{\mathrm{eff}} = 1/2$ bands toward half-filling while keeping the $J_{\mathrm{eff}} = 3/2$ bands away from it. Consequently, electron correlations induce an orbital-selective Mott transition in the $J_{\mathrm{eff}} = 1/2$ bands, whereas a band-insulating gap develops in the $J_{\mathrm{eff}} = 3/2$ bands, thereby stabilizing the unconventional insulating state in La$_3$Ir$_3$O$_{11}$. These findings provide new insights into the design and understanding of the insulating ground state of spin-orbit-coupled iridates.
\end{abstract}


\maketitle

%
%

Mott insulators, in which insulating behavior arises from strong electron-electron interactions rather than band filling, have been a cornerstone of modern condensed matter physics since their theoretical prediction~\cite{Imada1998}. They challenge the conventional band theory, as the insulating gap emerges from the Coulomb repulsion $U$ dominating over the kinetic energy $W$, leading to localized electrons. Doped Mott insulators, most notably the high-$T_c$ cuprates, have profoundly revolutionized our understanding of quantum materials by hosting intertwined orders, pseudogaps, and unconventional superconductivity~\cite{Dagotto1994,Lee2006}. In $5d$ transition metal oxides such as iridates, strong spin-orbit coupling (SOC) introduces a new dimension to Mott physics. In compounds like \SIO, the interplay between SOC and Coulomb interactions splits the $t_{2g}$ orbitals into spin-orbit-entangled $J_{\mathrm{eff}} = 1/2$ and $J_{\mathrm{eff}} = 3/2$ states, giving rise to a spin-orbit Mott insulator within the half-filled $J_{\mathrm{eff}} = 1/2$ band~\cite{Kim2009,Kim2008,Moon2009,Jackeli2009}. Analogous to cuprates, tuning this Mott state via chemical doping~\cite{Qi2012,Lee2012,Cao2016,Xu2020}, strain~\cite{Paris2020,Shrestha2022,Engstrom2023}, and dimensionality~\cite{Moon2008,Wang2013}, has revealed a rich set of emergent correlated phenomena, including novel charge and magnetic order\cite{Kim2014SC,Kim2014NC,Zhao2016,Zhou2017,Jeong2017,Wang2018,Cao2018,Murayama2021,Kim2024}, pseudogap features~\cite{Battisti2017,Seo2017,Louat2019}, and possible unconventional superconductivity~\cite{Wang2011,Yan2015}.

A critical question underlying these studies is whether a Mott insulating state can survive away from half filling. In conventional single-band Mott systems, as well as in the $J_{\mathrm{eff}} = 1/2$ Mott insulators, even weak doping rapidly suppresses the Mott gap and drives the system into a correlated metallic state~\cite{Imada1998,Dagotto1994,Lee2006,Qi2012,Lee2012,Cao2016,Xu2020}. In multi-orbital systems, additional orbital degrees of freedom enable more intricate correlation effects, such as orbital-selective Mott physics, as extensively discussed in iron-based superconductors and ruthenates~\cite{Medici2005,Medici2009,Medici2014,Yu2013PRL,Yu2017PRB,Kim2024PRL,Wang2004PRL,Neupane2009PRL}. Nevertheless, deviations from $3d^5$ or $4d^5$ fillings in these systems typically lead to Hund's metallic behavior. Consequently, stabilizing a correlated insulating state at non-integer filling remains highly non-trivial. Beyond orbital degrees of freedom, structural effects provide an alternative route to reshaping the electronic state in iridates. For example, lattice distortions of the IrO$_6$ octahedra can lift orbital degeneracies and renormalize bandwidths, while Ir-Ir dimerization induces strong bonding-antibonding splitting that reconstructs the low-energy electronic structure~\cite{Cao2018ROPP,Mazin2012PRL,Liu2012PRL,Gretarsson2013PRL,Dey2014PRB,Terzic2015,Hermann2018,Ye2018PRB,Wang2019,Zhao2019PRB,Khan2019PRB,Jeong2020PRL,Kumar2021PRB,Lane2020PRB}. Their interplay with strong spin-orbit coupling and electron correlations creates a highly entangled landscape in which insulating states may persist even at fractional fillings.

\begin{figure*}[tb]
\includegraphics[width=2\columnwidth]{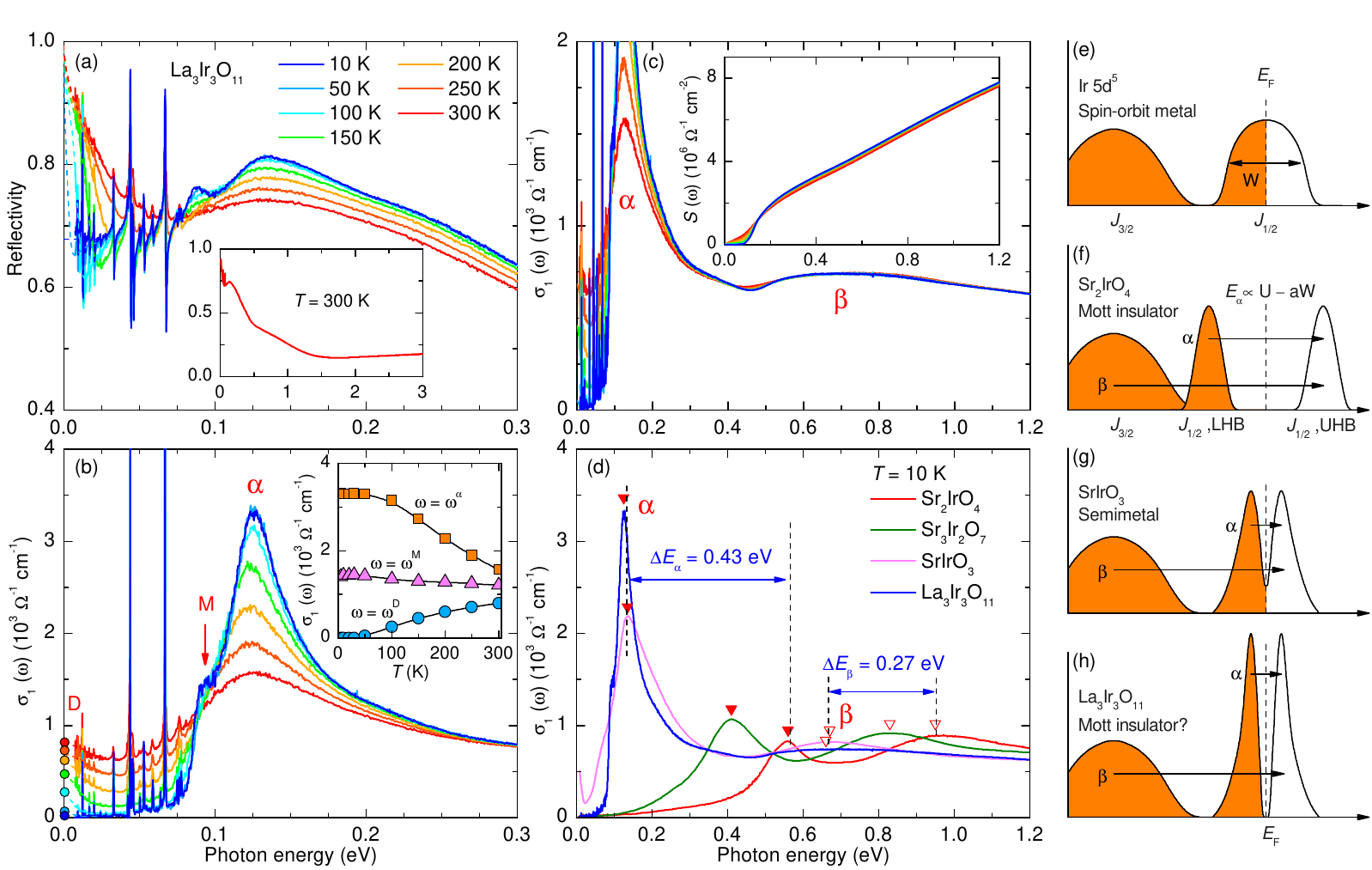}
\caption{ (color online) (a) Temperature-dependent reflectivity R($\omega$) of La$_3$Ir$_3$O$_{11}$ up to 0.3 eV. The dashed line represents the corresponding low-energy extrapolation. Inset: Spectrum up to 3 eV at 300 K. (b) Temperature-dependent optical conductivity up to 0.3 eV. Symbols on the $y$ axis denote dc conductivity values at corresponding temperatures from transport data. (c) Optical conductivity up to 1.2 eV to show the spectral changes at high frequencies. Inset: the energy dependence of spectral weight at several selected temperatures. (d) Comparison of $\sigma_1(\omega)$ between La$_3$Ir$_3$O$_{11}$ and the $5d^5$ iridates Sr$_2$IrO$_{4}$~\cite{Xu2020}, Sr$_3$Ir$_2$O$_{7}$~\cite{Seo2017}, and SrIrO$_{3}$~\cite{Fujioka2018} at 10~K. (e--h) Diagrams of density of states for different materials.}
\label{Fig1}
\end{figure*}

In this context, the geometrically frustrated iridate La$_3$Ir$_3$O$_{11}$ provides a particularly compelling platform. As shown in Fig.~\ref{Fig3}(g), it crystallizes in a cubic structure (space group $\mathrm{Pn}\bar{3}$), composed of a three-dimensional network of Ir$_{2}$O$_{10}$ dimers formed by pairs of edge-sharing IrO$_{6}$ octahedra. Importantly, all Ir ions are crystallographically equivalent and carry a uniform fractional valence of Ir$^{4.33+}$ (corresponding to a 5$d^{4.67}$ configuration) rather than a mixture of Ir$^{4+}$ and Ir$^{5+}$ states~\cite{Yang2019,Aoyama2019,Abraham1982}. This places La$_3$Ir$_3$O$_{11}$ effectively at $1/3$ hole doping relative to a 5$d^{5}$ spin-orbit Mott insulator, a regime where metallic behavior would typically be expected. Surprisingly, however, La$_3$Ir$_3$O$_{11}$ shows no evidence of coherent quasiparticle excitations, but instead remains insulating~\cite{Yang2019}, posing a direct challenge to conventional understanding of doped Mott systems.

In this Letter, we investigate the origin of this unconventional insulating state through optical conductivity measurements combined with first-principles calculations. We show that La$_3$Ir$_3$O$_{11}$ hosts a robust correlated insulating state, characterized by the collapse of the Drude response and the emergence of sharp low-energy excitations. Our theoretical analysis reveals that distortions within the Ir$_2$O$_{10}$ dimers split the $t_{2g}$ manifold in an orbital-selective manner, bringing the $J_{\mathrm{eff}} = 1/2$ bands close to half filling while pushing the $J_{\mathrm{eff}} = 3/2$ bands away. The resulting coexistence of a Mott gap in the $J_{\mathrm{eff}} = 1/2$ bands and a band gap in the $J_{\mathrm{eff}} = 3/2$ bands stabilizes a spin-orbit-assisted, orbital-selective insulating state at fractional filling in \LIO.

%

Sample synthesis, experimental methods, and details of Drude-Lorentz analysis and theoretical calculations are provided in the Supplemental Materials~\footnotemark[1].
%
%
%

Figure~\ref{Fig1}(a) displays the temperature-dependent reflectivity spectra of La$_3$Ir$_3$O$_{11}$ up to 0.3 eV. The inset of Fig.~\ref{Fig1}(a) shows the reflectivity at 300 K for the entire measured spectral range. At room temperature, the reflectivity exhibits a plasma edge below 0.05 eV, accompanied by a prominent bump feature around 0.13 eV. With decreasing temperature, the plasma edge shifts to lower energies and eventually disappears below 50 K, while the bump feature becomes more pronounced. In addition to these gross electronic features, a series of infrared-active phonon modes is present in the far-infrared region, which becomes sharper and more prominent at lower temperatures because of the reduced electron screening. Similar phenomena are frequently observed in the context of the metal-insulator transition in various correlated materials~\cite{Basov2005,Basov2011}. Overall, the significantly decreased or absent plasma edge, along with the prominent phonon behavior, points toward an insulating nature of electrons at low temperatures.

Figure~\ref{Fig1}(b) depicts the temperature dependence of the optical conductivity $\sigma_1(\omega)$ of La$_3$Ir$_3$O$_{11}$ up to 0.3 eV. The low-energy $\sigma_1(\omega)$ presents a Drude-like peak (labeled as D) with $\sigma_1(\omega \rightarrow 0) \simeq 800~\Omega^{-1}~\mathrm{cm}^{-1}$ at $T = 300$ K. Moving to higher frequencies, there is a broad absorption peak (labeled as $\alpha$) centered around 0.13 eV, consistent with the observation from the previous reflectivity spectrum. As the temperature decreases, the Drude-like peak gradually loses its spectral weight (SW), with $\sigma_1(\omega \rightarrow 0)$ decreasing in correlation with the inverse of the resistivity observed in transport. Meanwhile, the $\alpha$ peak gains more SW while maintaining a relatively unchanged peak position. Below $\sim$ 50 K, the Drude response is entirely quenched, resulting in an optical gap below approximately 0.06 eV. Notably, at low temperatures, a distinct shoulder feature (labeled as M) appears around 0.09 eV. The inset of Fig.~\ref{Fig1}(b) highlights the temperature evolution of the three features by showing the temperature dependence of $\sigma_1(\omega)$ at their peak positions.

Figure~\ref{Fig1}(c) shows the temperature-dependent $\sigma_1(\omega)$ up to 1.2 eV. In this range, the spectrum features a double peak structure, denoted as $\alpha$ and $\beta$. A direct comparison of $\sigma_1(\omega)$ for \LIO\ with the $5d^5$ iridates Sr$_2$IrO$_4$~\cite{Xu2020}, Sr$_3$Ir$_2$O$_7$~\cite{Seo2017}, and SrIrO$_3$~\cite{Fujioka2018} is presented in Fig.~\ref{Fig1}(d), all of which display similar features. In Sr$_2$IrO$_4$, as illustrated in Figs.~\ref{Fig1}(e--f), the on-site Coulomb interaction $U$ can split the half-filled $J_{\mathrm{eff}} = 1/2$ bands into a lower Hubbard band (LHB) and an upper Hubbard band (UHB), thus opening a Mott gap. In this context, the $\alpha$ peak arises from LHB $\rightarrow$ UHB transitions, while the $\beta$ peak originates from $J_{\mathrm{eff}}=3/2 \rightarrow$ UHB excitations~\cite{Kim2008,Moon2009,Zhang2013}. As shown in Fig.~\ref{Fig1}(d), both the $\alpha$ and $\beta$ peak positions ($E_{\alpha}$ and $E_{\beta}$) decrease systematically from Sr$_2$IrO$_4$ to SrIrO$_3$, reflecting increasing bandwidth ($W$)~\cite{Moon2008,Zhang2013}, since $\Delta \propto U - a W$, with $\Delta E_{\alpha}$ (0.43 eV) being roughly twice $\Delta E_{\beta}$ (0.27 eV), as sketched in Figs.~\ref{Fig1}(f--g). In \LIO, the $\alpha$ and $\beta$ peaks nearly coincide with those of SrIrO$_3$, implying comparable $J_{\mathrm{eff}}=1/2$ bandwidths. However, unlike the semimetallic SrIrO$_3$, \LIO\ develops an optical gap, and its $\alpha$ peak is strikingly sharper, indicating low-energy excitations associated with quasi-flat bands stabilized by strong electron correlations. These observations suggest that the $\alpha$ and $\beta$ features in \LIO\ share the same underlying physics as in \SIO, as sketched in Figs.~\ref{Fig1}(f) and \ref{Fig1}(h), with the robust Mott correlations underpinning its insulating state in \LIO.

\begin{figure}[tb]
\includegraphics[width=\columnwidth]{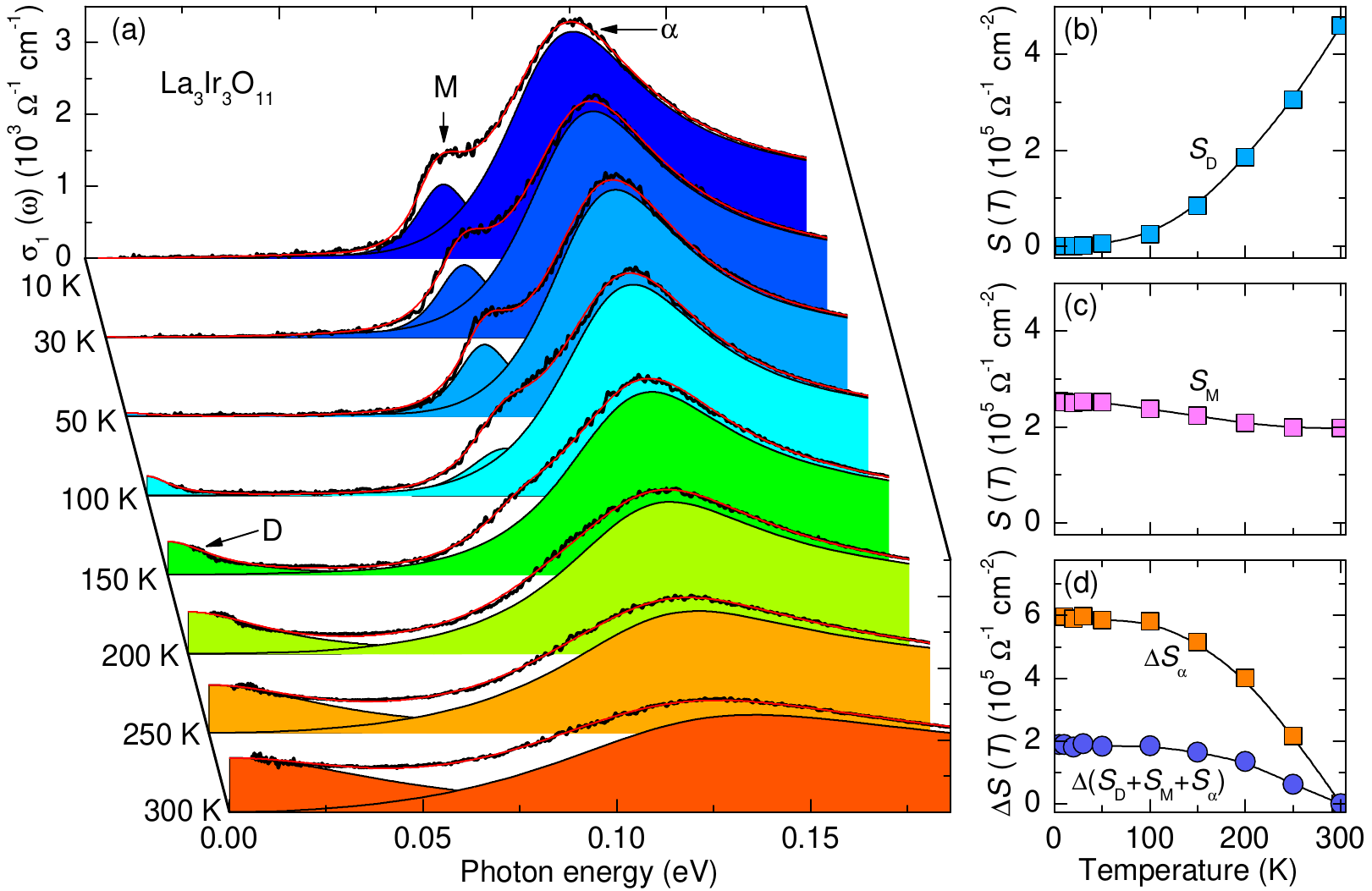}
\caption{ (color online) (a) Drude-Lorentz fits applied to $\sigma_1(\omega)$ spectra. (b--d), Temperature dependence of spectral weight for different components. $\Delta S(T) = S(T) - S(300~\mathrm{K})$ represents the difference in spectral weight relative to the value at 300 K.}
\label{Fig2}
\end{figure}

The role of Mott correlations is further evident by the SW analysis, quantified by $S(\omega,T) = \int_{0}^{\omega} \sigma_{1}(\omega,T)d\omega$. This analysis allows us to identify the nature of the insulating state, ``like band insulator or Mott insulator''~\cite{Phillips2010}. The energy dependence of $S(\omega,T)$ at several representative temperatures is showed in the inset of Fig.~\ref{Fig1}(c). With decreasing temperature, the free carrier response is completely quenched at low temperatures, while the lost SW does not fully recover within the $\alpha$ peak energy scale. Instead, it redistributes over a broad energy range, extending beyond 1.5 eV, which is comparable with the energy associated with the local on-site Coulomb interaction $U$ of Ir $5d$ orbitals~\cite{Dirk1988}. Thus, the observed anomalous SW transfer can be attributed to the ``Mottness'' of this system in \LIO.

\begin{figure*}[tb]
\includegraphics[width=2\columnwidth]{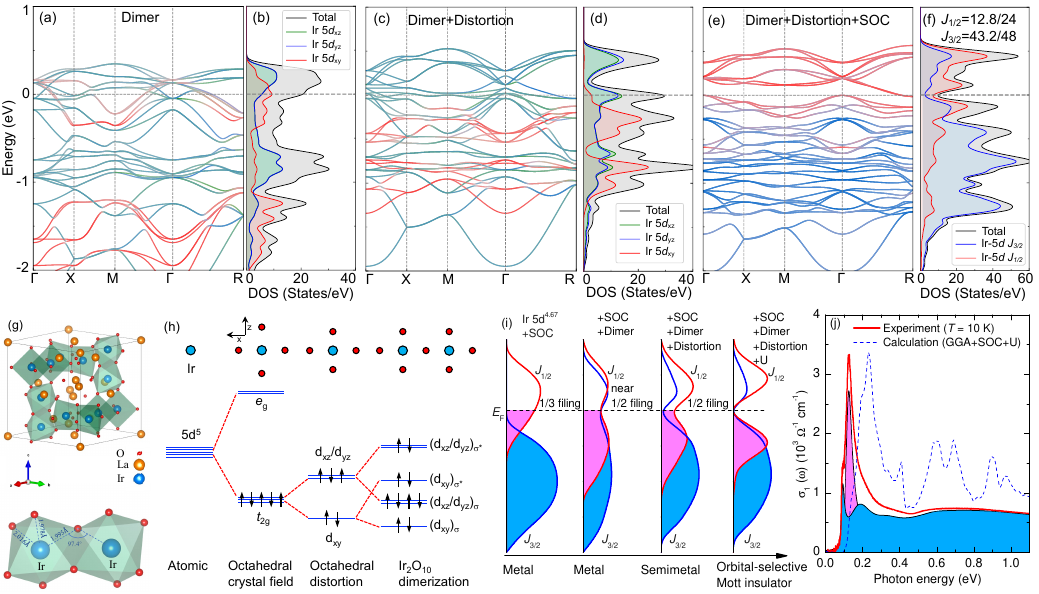}
\caption{ (color online) Tight-binding band structure and orbital-resolved density of states (DOS) for \LIO\ with (a,b) dimerization only, (c,d) dimerization and octahedral distortion, and (e,f) dimerization, distortion, and spin-orbit coupling (SOC). (g) Crystal structure of \LIO\ and the Ir$_2$O$_{10}$ dimer structure. (h) Orbital-energy-level diagram for the IrO$_6$ octahedra under various conditions. (i) DOS schematics within the $J_{\mathrm{eff}}$ framework for \LIO\ under the corresponding conditions. (j) Comparison between the experimental and calculated optical conductivity $\sigma_1(\omega)$.}
\label{Fig3}
\end{figure*}
To better illustrate the temperature evolution of various components in the $\sigma_1(\omega)$ spectra, we employ the Drude-Lorentz model to fit the data (see details in the Supplemental Materials~\footnotemark[1]). As shown in Fig~\ref{Fig2}(a), after subtracting the phonon modes, the low-energy $\sigma_1(\omega)$ can be decomposed into three components: the Drude response (D), the shoulder feature (M), and the $\alpha$ excitations. Figs.~\ref{Fig2}(b-d) show the temperature dependence of SW for each component. Obviously, the SW of the M and $\alpha$ components increases with decreasing temperature, while the SW of the Drude response diminishes. Moreover, the accumulated SW of the M and $\alpha$ excitations exceeds the SW lost by the Drude response, indicating that additional SW that is transferred from higher energies, which also aligns with our previous analysis.

Based on the above experimental results and discussions, we establish that the insulating state and the sharp $\alpha$ peak arise from the excitations across Mott-Hubbard bands. Nevertheless, two key puzzles immediately arise. First, how can \LIO\ maintain a robust Mott insulating ground state at a fractional filling of Ir $5d^{4.67}$, which naturally favors a metallic state? Second, what is the origin of the shoulder feature (M) observed within the Mott gap? These issues indicate that the correlated insulating state in \LIO\ is unconventional and cannot be fully explained by a single-band Mott-Hubbard framework. Regarding the origin of the M mode, several possibilities can be excluded. In-gap states are unlikely, as such excitations typically gain SW from the main Mott excitations (the $\alpha$ peak), as observed in \SIROx~\cite{Xu2020}, in clear contradiction with our experimental observations in \LIO. A magnetic origin, such as exciton and magnon sideband absorptions observed in rare-earth iridates~\cite{Ueda2016PRB}, is also improbable, given the absence of long-range magnetic order in La$_3$Ir$_{3}$O$_{11}$~\cite{Yang2019}.

To further elucidate the origin of the insulating state and low-energy excitations in \LIO, we performed systematic band calculations under various conditions (see Section C of the Supplemental Materials~\footnotemark[1]). These calculations allow us to disentangle the distinct roles of dimerization, octahedral distortion, spin-orbit coupling, and electron correlations, and to clarify how their cooperative interplay stabilizes the observed insulating state. We first examine the structural effects. In the absence of SOC and electron correlations, as shown in Figs.~\ref{Fig3}(a--d), the system exhibits a metallic electronic structure, with states near the Fermi level dominated by Ir $t_{2g}$ orbitals due to the octahedral crystal-field splitting. The formation of Ir$_2$O$_{10}$ dimers gives rise to a pronounced bonding-antibonding splitting of the $t_{2g}$ orbitals [Figs.~\ref{Fig3}(a--b)], reflecting the molecular-orbital character imposed by the dimerized units. When the experimentally observed octahedral distortion (compressed along the $z$ axis) is included [Figs.~\ref{Fig3}(c--d)], the $t_{2g}$ degeneracy is further lifted: the $d_{xz}/d_{yz}$ orbitals are shifted to higher energies and become less occupied, whereas the $d_{xy}$ orbital is pushed to lower energies and becomes more occupied. This progressive orbital reconstruction, driven by crystal-field, octahedral distortions, and dimerization, is schematically summarized in Fig.~\ref{Fig3}(h).

When SOC is explicitly included, the system evolves into a compensated semimetal, featuring hole pockets at the $\Gamma$ and electron pockets at R, as shown in Figs.~\ref{Fig3}(e--f). Analysis in the $J_{\mathrm{eff}}$ basis reveals two consequences promoted by octahedral distortion, as compared in Fig.~S6 in the Supplemental Materials~\footnotemark[1]. First, the separation between the $J_{\mathrm{eff}} = 1/2$ and $J_{\mathrm{eff}} = 3/2$ manifolds is significantly enhanced. Second the $J_{\mathrm{eff}} = 1/2$ bands are driven close to half filling (12.8/24 electrons). The near-half-filled condition naturally amplifies the susceptibility of the $J_{\mathrm{eff}}=1/2$ bands to an orbital-selective Mott transition. At the same time, the distortion opens a small semiconducting gap within the $J_{\mathrm{eff}}=3/2$ bands. Within this $J_\mathrm{eff}$ framework, the electronic evolution of \LIO\ can be summarized schematically in Fig.~\ref{Fig3}(i): dimerization establishes the basic bonding-antibonding structure and suppresses the density of states near the Fermi level; octahedral distortion enhances orbital differentiation and drives the $J_{\mathrm{eff}} = 1/2$ bands toward half filling; electron correlations then selectively act on this nearly half-filled sector to open a Mott gap. Indeed, when correlations are included via GGA+SOC+$U$ calculations (Fig.~S7 in the Supplemental Materials~\footnotemark[1]), the residual density of states at the Fermi level is removed and a global insulating gap opens. The resulting ground state is therefore characterized by a Mott gap in the $J_{\mathrm{eff}}=1/2$ bands coexisting with a band gap in the $J_{\mathrm{eff}}=3/2$ bands, stabilizing the unconventional insulating state in \LIO.

This dual-gap electronic structure naturally accounts for the optical response. As shown in Fig.~\ref{Fig3}(j), the calculated optical conductivity reproduces the main experimental features, albeit with broader peaks and a systematic blueshift due to the underestimation of correlation-induced bandwidth renormalization in DFT+$U$ calculations~\cite{Zhang2013}. Within coexistence of Mott and band gaps, the low-energy optical response of \LIO\ can be qualitatively decomposed into two components: (i) Mott-gap excitations of the $J_{\mathrm{eff}} = 1/2$ bands, which give rise to the sharp $\alpha$ peak (magenta shading), and (ii) interband excitations across the gap in the $J_{\mathrm{eff}} = 3/2$ bands, responsible for the shoulder (M mode) and the broad background beneath the $\alpha$ peak (light-blue shading).

%
%
Finally, we discuss the implications of our work. At the materials level, the relatively large bandwidth in \LIO\ places the ratio $U/W$ close to the critical value for a Mott transition, making the insulating state highly sensitive to external parameters such as temperature. As a result, pronounced SW transfer between the Mott gap and Drude-like excitations is observed. In addition, variations in sample growth conditions can effectively tune lattice distortion or carrier filling, potentially driving the system back into a semimetallic state, as reported in previous studies in {\LIO}~\cite{Aoyama2019}. Such high tunability establishes \LIO\ as a promising platform for controlling correlated quantum phases. More fundamentally, our findings extend the paradigm of Mott insulating behavior in spin-orbit-coupled systems beyond half-filled single-band models. We demonstrate that orbital-selective correlations can stabilize a Mott-insulating $J_{\mathrm{eff}}=1/2$ band coexisting with a band-insulating $J_{\mathrm{eff}}=3/2$ background, even at fractional valence. The cooperative interplay of structural distortions, SOC, and multiorbital correlations thus provides a general route to unconventional insulating states. In this context, our work is consistent with recent theoretical proposals of orbital-selective Mott physics in spin-orbit-coupled multiorbital systems~\cite{Chen2025,Li2025}, and suggests new strategies for engineering novel quantum phases in correlated materials.

%
%
\begin{acknowledgments}
We acknowledge valuable discussions with Zhenyu Zhang, Yuanyao He, Changming Yue, and Sen Zhou. We also thank Yuyan Han for conducting the DC transport measurements at High Magnetic Field Laboratory, Hefei Institutes of Physical Science, Chinese Academy of Sciences. This work was supported by the National Natural Science Foundation of China (Grant Nos. 12274442 and 12474239), the National Key Research and Development Program of China (Grants Nos. 2022YFA1403901, 2024YFA1408301, and 2023YFA1406002), and the Special Support Program for Exceptional Talents of Henan University  (Grant No.CX3050A0980175 ). J.Y. acknowledges support by China Postdoctoral Science Foundation (Grant No. 2022M712845).
\end{acknowledgments}

%

%


\begin{thebibliography}{75}%
\makeatletter
\providecommand \@ifxundefined [1]{%
 \@ifx{#1\undefined}
}%
\providecommand \@ifnum [1]{%
 \ifnum #1\expandafter \@firstoftwo
 \else \expandafter \@secondoftwo
 \fi
}%
\providecommand \@ifx [1]{%
 \ifx #1\expandafter \@firstoftwo
 \else \expandafter \@secondoftwo
 \fi
}%
\providecommand \natexlab [1]{#1}%
\providecommand \enquote  [1]{``#1''}%
\providecommand \bibnamefont  [1]{#1}%
\providecommand \bibfnamefont [1]{#1}%
\providecommand \citenamefont [1]{#1}%
\providecommand \href@noop [0]{\@secondoftwo}%
\providecommand \href [0]{\begingroup \@sanitize@url \@href}%
\providecommand \@href[1]{\@@startlink{#1}\@@href}%
\providecommand \@@href[1]{\endgroup#1\@@endlink}%
\providecommand \@sanitize@url [0]{\catcode `\\12\catcode `\$12\catcode
  `\&12\catcode `\#12\catcode `\^12\catcode `\_12\catcode `\%12\relax}%
\providecommand \@@startlink[1]{}%
\providecommand \@@endlink[0]{}%
\providecommand \url  [0]{\begingroup\@sanitize@url \@url }%
\providecommand \@url [1]{\endgroup\@href {#1}{\urlprefix }}%
\providecommand \urlprefix  [0]{URL }%
\providecommand \Eprint [0]{\href }%
\providecommand \doibase [0]{https://doi.org/}%
\providecommand \selectlanguage [0]{\@gobble}%
\providecommand \bibinfo  [0]{\@secondoftwo}%
\providecommand \bibfield  [0]{\@secondoftwo}%
\providecommand \translation [1]{[#1]}%
\providecommand \BibitemOpen [0]{}%
\providecommand \bibitemStop [0]{}%
\providecommand \bibitemNoStop [0]{.\EOS\space}%
\providecommand \EOS [0]{\spacefactor3000\relax}%
\providecommand \BibitemShut  [1]{\csname bibitem#1\endcsname}%
\let\auto@bib@innerbib\@empty
\bibitem [{\citenamefont {Imada}\ \emph {et~al.}(1998)\citenamefont {Imada},
  \citenamefont {Fujimori},\ and\ \citenamefont {Tokura}}]{Imada1998}%
  \BibitemOpen
  \bibfield  {author} {\bibinfo {author} {\bibfnamefont {M.}~\bibnamefont
  {Imada}}, \bibinfo {author} {\bibfnamefont {A.}~\bibnamefont {Fujimori}},\
  and\ \bibinfo {author} {\bibfnamefont {Y.}~\bibnamefont {Tokura}},\
  }\bibfield  {title} {\bibinfo {title} {{Metal-insulator transitions}},\
  }\href {https://doi.org/10.1103/RevModPhys.70.1039} {\bibfield  {journal}
  {\bibinfo  {journal} {Rev. Mod. Phys.}\ }\textbf {\bibinfo {volume} {70}},\
  \bibinfo {pages} {1039} (\bibinfo {year} {1998})}\BibitemShut {NoStop}%
\bibitem [{\citenamefont {Dagotto}(1994)}]{Dagotto1994}%
  \BibitemOpen
  \bibfield  {author} {\bibinfo {author} {\bibfnamefont {E.}~\bibnamefont
  {Dagotto}},\ }\bibfield  {title} {\bibinfo {title} {{Correlated electrons in
  high-temperature superconductors}},\ }\href
  {https://doi.org/10.1103/RevModPhys.66.763} {\bibfield  {journal} {\bibinfo
  {journal} {Rev. Mod. Phys.}\ }\textbf {\bibinfo {volume} {66}},\ \bibinfo
  {pages} {763} (\bibinfo {year} {1994})}\BibitemShut {NoStop}%
\bibitem [{\citenamefont {Lee}\ \emph {et~al.}(2006)\citenamefont {Lee},
  \citenamefont {Nagaosa},\ and\ \citenamefont {Wen}}]{Lee2006}%
  \BibitemOpen
  \bibfield  {author} {\bibinfo {author} {\bibfnamefont {P.~A.}\ \bibnamefont
  {Lee}}, \bibinfo {author} {\bibfnamefont {N.}~\bibnamefont {Nagaosa}},\ and\
  \bibinfo {author} {\bibfnamefont {X.-G.}\ \bibnamefont {Wen}},\ }\bibfield
  {title} {\bibinfo {title} {Doping a mott insulator: Physics of
  high-temperature superconductivity},\ }\href
  {https://doi.org/10.1103/RevModPhys.78.17} {\bibfield  {journal} {\bibinfo
  {journal} {Rev. Mod. Phys.}\ }\textbf {\bibinfo {volume} {78}},\ \bibinfo
  {pages} {17} (\bibinfo {year} {2006})}\BibitemShut {NoStop}%
\bibitem [{\citenamefont {Kim}\ \emph {et~al.}(2009)\citenamefont {Kim},
  \citenamefont {Ohsumi}, \citenamefont {Komesu}, \citenamefont {Sakai},
  \citenamefont {Morita}, \citenamefont {Takagi},\ and\ \citenamefont
  {Arima}}]{Kim2009}%
  \BibitemOpen
  \bibfield  {author} {\bibinfo {author} {\bibfnamefont {B.~J.}\ \bibnamefont
  {Kim}}, \bibinfo {author} {\bibfnamefont {H.}~\bibnamefont {Ohsumi}},
  \bibinfo {author} {\bibfnamefont {T.}~\bibnamefont {Komesu}}, \bibinfo
  {author} {\bibfnamefont {S.}~\bibnamefont {Sakai}}, \bibinfo {author}
  {\bibfnamefont {T.}~\bibnamefont {Morita}}, \bibinfo {author} {\bibfnamefont
  {H.}~\bibnamefont {Takagi}},\ and\ \bibinfo {author} {\bibfnamefont
  {T.}~\bibnamefont {Arima}},\ }\bibfield  {title} {\bibinfo {title}
  {{Phase-Sensitive Observation of a Spin-Orbital Mott State in
  ${\mathrm{Sr}}_{2}{\mathrm{IrO}}_{4}$}},\ }\href
  {https://doi.org/10.1126/science.1167106} {\bibfield  {journal} {\bibinfo
  {journal} {Science}\ }\textbf {\bibinfo {volume} {323}},\ \bibinfo {pages}
  {1329} (\bibinfo {year} {2009})}\BibitemShut {NoStop}%
\bibitem [{\citenamefont {Kim}\ \emph {et~al.}(2008)\citenamefont {Kim},
  \citenamefont {Jin}, \citenamefont {Moon}, \citenamefont {Kim}, \citenamefont
  {Park}, \citenamefont {Leem}, \citenamefont {Yu}, \citenamefont {Noh},
  \citenamefont {Kim}, \citenamefont {Oh}, \citenamefont {Park}, \citenamefont
  {Durairaj}, \citenamefont {Cao},\ and\ \citenamefont {Rotenberg}}]{Kim2008}%
  \BibitemOpen
  \bibfield  {author} {\bibinfo {author} {\bibfnamefont {B.~J.}\ \bibnamefont
  {Kim}}, \bibinfo {author} {\bibfnamefont {H.}~\bibnamefont {Jin}}, \bibinfo
  {author} {\bibfnamefont {S.~J.}\ \bibnamefont {Moon}}, \bibinfo {author}
  {\bibfnamefont {J.-Y.}\ \bibnamefont {Kim}}, \bibinfo {author} {\bibfnamefont
  {B.-G.}\ \bibnamefont {Park}}, \bibinfo {author} {\bibfnamefont {C.~S.}\
  \bibnamefont {Leem}}, \bibinfo {author} {\bibfnamefont {J.}~\bibnamefont
  {Yu}}, \bibinfo {author} {\bibfnamefont {T.~W.}\ \bibnamefont {Noh}},
  \bibinfo {author} {\bibfnamefont {C.}~\bibnamefont {Kim}}, \bibinfo {author}
  {\bibfnamefont {S.-J.}\ \bibnamefont {Oh}}, \bibinfo {author} {\bibfnamefont
  {J.-H.}\ \bibnamefont {Park}}, \bibinfo {author} {\bibfnamefont
  {V.}~\bibnamefont {Durairaj}}, \bibinfo {author} {\bibfnamefont
  {G.}~\bibnamefont {Cao}},\ and\ \bibinfo {author} {\bibfnamefont
  {E.}~\bibnamefont {Rotenberg}},\ }\bibfield  {title} {\bibinfo {title}
  {{Novel ${J}_{\mathrm{eff}}=1/2$ Mott State Induced by Relativistic
  Spin-Orbit Coupling in ${\mathrm{Sr}}_{2}{\mathrm{IrO}}_{4}$}},\ }\href
  {https://doi.org/10.1103/PhysRevLett.101.076402} {\bibfield  {journal}
  {\bibinfo  {journal} {Phys. Rev. Lett.}\ }\textbf {\bibinfo {volume} {101}},\
  \bibinfo {pages} {076402} (\bibinfo {year} {2008})}\BibitemShut {NoStop}%
\bibitem [{\citenamefont {Moon}\ \emph {et~al.}(2009)\citenamefont {Moon},
  \citenamefont {Jin}, \citenamefont {Choi}, \citenamefont {Lee}, \citenamefont
  {Seo}, \citenamefont {Yu}, \citenamefont {Cao}, \citenamefont {Noh},\ and\
  \citenamefont {Lee}}]{Moon2009}%
  \BibitemOpen
  \bibfield  {author} {\bibinfo {author} {\bibfnamefont {S.~J.}\ \bibnamefont
  {Moon}}, \bibinfo {author} {\bibfnamefont {H.}~\bibnamefont {Jin}}, \bibinfo
  {author} {\bibfnamefont {W.~S.}\ \bibnamefont {Choi}}, \bibinfo {author}
  {\bibfnamefont {J.~S.}\ \bibnamefont {Lee}}, \bibinfo {author} {\bibfnamefont
  {S.~S.~A.}\ \bibnamefont {Seo}}, \bibinfo {author} {\bibfnamefont
  {J.}~\bibnamefont {Yu}}, \bibinfo {author} {\bibfnamefont {G.}~\bibnamefont
  {Cao}}, \bibinfo {author} {\bibfnamefont {T.~W.}\ \bibnamefont {Noh}},\ and\
  \bibinfo {author} {\bibfnamefont {Y.~S.}\ \bibnamefont {Lee}},\ }\bibfield
  {title} {\bibinfo {title} {{Temperature dependence of the electronic
  structure of the ${J}_{\text{eff}}=\frac{1}{2}$ Mott insulator
  ${\text{Sr}}_{2}{\text{IrO}}_{4}$ studied by optical spectroscopy}},\ }\href
  {https://doi.org/10.1103/PhysRevB.80.195110} {\bibfield  {journal} {\bibinfo
  {journal} {Phys. Rev. B}\ }\textbf {\bibinfo {volume} {80}},\ \bibinfo
  {pages} {195110} (\bibinfo {year} {2009})}\BibitemShut {NoStop}%
\bibitem [{\citenamefont {Jackeli}\ and\ \citenamefont
  {Khaliullin}(2009)}]{Jackeli2009}%
  \BibitemOpen
  \bibfield  {author} {\bibinfo {author} {\bibfnamefont {G.}~\bibnamefont
  {Jackeli}}\ and\ \bibinfo {author} {\bibfnamefont {G.}~\bibnamefont
  {Khaliullin}},\ }\bibfield  {title} {\bibinfo {title} {{Mott Insulators in
  the Strong Spin-Orbit Coupling Limit: From Heisenberg to a Quantum Compass
  and Kitaev Models}},\ }\href {https://doi.org/10.1103/PhysRevLett.102.017205}
  {\bibfield  {journal} {\bibinfo  {journal} {Phys. Rev. Lett.}\ }\textbf
  {\bibinfo {volume} {102}},\ \bibinfo {pages} {017205} (\bibinfo {year}
  {2009})}\BibitemShut {NoStop}%
\bibitem [{\citenamefont {Qi}\ \emph {et~al.}(2012)\citenamefont {Qi},
  \citenamefont {Korneta}, \citenamefont {Li}, \citenamefont {Butrouna},
  \citenamefont {Cao}, \citenamefont {Wan}, \citenamefont {Schlottmann},
  \citenamefont {Kaul},\ and\ \citenamefont {Cao}}]{Qi2012}%
  \BibitemOpen
  \bibfield  {author} {\bibinfo {author} {\bibfnamefont {T.~F.}\ \bibnamefont
  {Qi}}, \bibinfo {author} {\bibfnamefont {O.~B.}\ \bibnamefont {Korneta}},
  \bibinfo {author} {\bibfnamefont {L.}~\bibnamefont {Li}}, \bibinfo {author}
  {\bibfnamefont {K.}~\bibnamefont {Butrouna}}, \bibinfo {author}
  {\bibfnamefont {V.~S.}\ \bibnamefont {Cao}}, \bibinfo {author} {\bibfnamefont
  {X.}~\bibnamefont {Wan}}, \bibinfo {author} {\bibfnamefont {P.}~\bibnamefont
  {Schlottmann}}, \bibinfo {author} {\bibfnamefont {R.~K.}\ \bibnamefont
  {Kaul}},\ and\ \bibinfo {author} {\bibfnamefont {G.}~\bibnamefont {Cao}},\
  }\bibfield  {title} {\bibinfo {title} {{Spin-orbit tuned metal-insulator
  transitions in single-crystal
  Sr${}_{2}$Ir${}_{1\ensuremath{-}x}$Rh${}_{x}$O${}_{4}$
  ($0\ensuremath{\le}x\ensuremath{\le}1$)}},\ }\href
  {https://doi.org/10.1103/PhysRevB.86.125105} {\bibfield  {journal} {\bibinfo
  {journal} {Phys. Rev. B}\ }\textbf {\bibinfo {volume} {86}},\ \bibinfo
  {pages} {125105} (\bibinfo {year} {2012})}\BibitemShut {NoStop}%
\bibitem [{\citenamefont {Lee}\ \emph {et~al.}(2012)\citenamefont {Lee},
  \citenamefont {Krockenberger}, \citenamefont {Takahashi}, \citenamefont
  {Kawasaki},\ and\ \citenamefont {Tokura}}]{Lee2012}%
  \BibitemOpen
  \bibfield  {author} {\bibinfo {author} {\bibfnamefont {J.~S.}\ \bibnamefont
  {Lee}}, \bibinfo {author} {\bibfnamefont {Y.}~\bibnamefont {Krockenberger}},
  \bibinfo {author} {\bibfnamefont {K.~S.}\ \bibnamefont {Takahashi}}, \bibinfo
  {author} {\bibfnamefont {M.}~\bibnamefont {Kawasaki}},\ and\ \bibinfo
  {author} {\bibfnamefont {Y.}~\bibnamefont {Tokura}},\ }\bibfield  {title}
  {\bibinfo {title} {{Insulator-metal transition driven by change of doping and
  spin-orbit interaction in Sr${}_{2}$IrO${}_{4}$}},\ }\href
  {https://doi.org/10.1103/PhysRevB.85.035101} {\bibfield  {journal} {\bibinfo
  {journal} {Phys. Rev. B}\ }\textbf {\bibinfo {volume} {85}},\ \bibinfo
  {pages} {035101} (\bibinfo {year} {2012})}\BibitemShut {NoStop}%
\bibitem [{\citenamefont {Cao}\ \emph {et~al.}(2016)\citenamefont {Cao},
  \citenamefont {Wang}, \citenamefont {Waugh}, \citenamefont {Reber},
  \citenamefont {Li}, \citenamefont {Zhou}, \citenamefont {Parham},
  \citenamefont {Park}, \citenamefont {Plumb}, \citenamefont {Rotenberg},
  \citenamefont {Bostwick}, \citenamefont {Denlinger}, \citenamefont {Qi},
  \citenamefont {Hermele}, \citenamefont {Cao},\ and\ \citenamefont
  {Dessau}}]{Cao2016}%
  \BibitemOpen
  \bibfield  {author} {\bibinfo {author} {\bibfnamefont {Y.}~\bibnamefont
  {Cao}}, \bibinfo {author} {\bibfnamefont {Q.}~\bibnamefont {Wang}}, \bibinfo
  {author} {\bibfnamefont {J.~A.}\ \bibnamefont {Waugh}}, \bibinfo {author}
  {\bibfnamefont {T.~J.}\ \bibnamefont {Reber}}, \bibinfo {author}
  {\bibfnamefont {H.}~\bibnamefont {Li}}, \bibinfo {author} {\bibfnamefont
  {X.}~\bibnamefont {Zhou}}, \bibinfo {author} {\bibfnamefont {S.}~\bibnamefont
  {Parham}}, \bibinfo {author} {\bibfnamefont {S.-R.}\ \bibnamefont {Park}},
  \bibinfo {author} {\bibfnamefont {N.~C.}\ \bibnamefont {Plumb}}, \bibinfo
  {author} {\bibfnamefont {E.}~\bibnamefont {Rotenberg}}, \bibinfo {author}
  {\bibfnamefont {A.}~\bibnamefont {Bostwick}}, \bibinfo {author}
  {\bibfnamefont {J.~D.}\ \bibnamefont {Denlinger}}, \bibinfo {author}
  {\bibfnamefont {T.}~\bibnamefont {Qi}}, \bibinfo {author} {\bibfnamefont
  {M.~A.}\ \bibnamefont {Hermele}}, \bibinfo {author} {\bibfnamefont
  {G.}~\bibnamefont {Cao}},\ and\ \bibinfo {author} {\bibfnamefont {D.~S.}\
  \bibnamefont {Dessau}},\ }\bibfield  {title} {\bibinfo {title} {{Hallmarks of
  the Mott-metal crossover in the hole-doped pseudospin-1/2 Mott insulator
  Sr$_2$IrO$_4$}},\ }\href {https://doi.org/10.1038/ncomms11367} {\bibfield
  {journal} {\bibinfo  {journal} {Nat. Commun.}\ }\textbf {\bibinfo {volume}
  {7}},\ \bibinfo {pages} {11367} (\bibinfo {year} {2016})}\BibitemShut
  {NoStop}%
\bibitem [{\citenamefont {Xu}\ \emph {et~al.}(2020)\citenamefont {Xu},
  \citenamefont {Marsik}, \citenamefont {Sheveleva}, \citenamefont {Lyzwa},
  \citenamefont {Louat}, \citenamefont {Brouet}, \citenamefont {Munzar},\ and\
  \citenamefont {Bernhard}}]{Xu2020}%
  \BibitemOpen
  \bibfield  {author} {\bibinfo {author} {\bibfnamefont {B.}~\bibnamefont
  {Xu}}, \bibinfo {author} {\bibfnamefont {P.}~\bibnamefont {Marsik}}, \bibinfo
  {author} {\bibfnamefont {E.}~\bibnamefont {Sheveleva}}, \bibinfo {author}
  {\bibfnamefont {F.}~\bibnamefont {Lyzwa}}, \bibinfo {author} {\bibfnamefont
  {A.}~\bibnamefont {Louat}}, \bibinfo {author} {\bibfnamefont
  {V.}~\bibnamefont {Brouet}}, \bibinfo {author} {\bibfnamefont
  {D.}~\bibnamefont {Munzar}},\ and\ \bibinfo {author} {\bibfnamefont
  {C.}~\bibnamefont {Bernhard}},\ }\bibfield  {title} {\bibinfo {title}
  {{Optical Signature of a Crossover from Mott- to Slater-Type Gap in
  ${\mathrm{Sr}}_{2}{\mathrm{Ir}}_{1\ensuremath{-}x}{\mathrm{Rh}}_{x}{\mathrm{O}}_{4}$}},\
  }\href {https://doi.org/10.1103/PhysRevLett.124.027402} {\bibfield  {journal}
  {\bibinfo  {journal} {Phys. Rev. Lett.}\ }\textbf {\bibinfo {volume} {124}},\
  \bibinfo {pages} {027402} (\bibinfo {year} {2020})}\BibitemShut {NoStop}%
\bibitem [{\citenamefont {Paris}\ \emph {et~al.}(2020)\citenamefont {Paris},
  \citenamefont {Tseng}, \citenamefont {P{\"{a}}rschke}, \citenamefont {Zhang},
  \citenamefont {Upton}, \citenamefont {Efimenko}, \citenamefont {Rolfs},
  \citenamefont {McNally}, \citenamefont {Maurel}, \citenamefont {Naamneh},
  \citenamefont {Caputo}, \citenamefont {Strocov}, \citenamefont {Wang},
  \citenamefont {Casa}, \citenamefont {Schneider}, \citenamefont
  {Pomjakushina}, \citenamefont {Wohlfeld}, \citenamefont {Radovic},\ and\
  \citenamefont {Schmitt}}]{Paris2020}%
  \BibitemOpen
  \bibfield  {author} {\bibinfo {author} {\bibfnamefont {E.}~\bibnamefont
  {Paris}}, \bibinfo {author} {\bibfnamefont {Y.}~\bibnamefont {Tseng}},
  \bibinfo {author} {\bibfnamefont {E.~M.}\ \bibnamefont {P{\"{a}}rschke}},
  \bibinfo {author} {\bibfnamefont {W.}~\bibnamefont {Zhang}}, \bibinfo
  {author} {\bibfnamefont {M.~H.}\ \bibnamefont {Upton}}, \bibinfo {author}
  {\bibfnamefont {A.}~\bibnamefont {Efimenko}}, \bibinfo {author}
  {\bibfnamefont {K.}~\bibnamefont {Rolfs}}, \bibinfo {author} {\bibfnamefont
  {D.~E.}\ \bibnamefont {McNally}}, \bibinfo {author} {\bibfnamefont
  {L.}~\bibnamefont {Maurel}}, \bibinfo {author} {\bibfnamefont
  {M.}~\bibnamefont {Naamneh}}, \bibinfo {author} {\bibfnamefont
  {M.}~\bibnamefont {Caputo}}, \bibinfo {author} {\bibfnamefont {V.~N.}\
  \bibnamefont {Strocov}}, \bibinfo {author} {\bibfnamefont {Z.}~\bibnamefont
  {Wang}}, \bibinfo {author} {\bibfnamefont {D.}~\bibnamefont {Casa}}, \bibinfo
  {author} {\bibfnamefont {C.~W.}\ \bibnamefont {Schneider}}, \bibinfo {author}
  {\bibfnamefont {E.}~\bibnamefont {Pomjakushina}}, \bibinfo {author}
  {\bibfnamefont {K.}~\bibnamefont {Wohlfeld}}, \bibinfo {author}
  {\bibfnamefont {M.}~\bibnamefont {Radovic}},\ and\ \bibinfo {author}
  {\bibfnamefont {T.}~\bibnamefont {Schmitt}},\ }\bibfield  {title} {\bibinfo
  {title} {{Strain engineering of the charge and spin-orbital interactions in
  Sr$_2$IrO$_4$}},\ }\href {https://doi.org/10.1073/pnas.2012043117} {\bibfield
   {journal} {\bibinfo  {journal} {Proceedings of the National Academy of
  Sciences}\ }\textbf {\bibinfo {volume} {117}},\ \bibinfo {pages} {24764}
  (\bibinfo {year} {2020})}\BibitemShut {NoStop}%
\bibitem [{\citenamefont {Shrestha}\ \emph {et~al.}(2022)\citenamefont
  {Shrestha}, \citenamefont {Krautloher}, \citenamefont {Zhu}, \citenamefont
  {Kim}, \citenamefont {Hwang}, \citenamefont {Kim}, \citenamefont {Kim},
  \citenamefont {Keimer},\ and\ \citenamefont {Seo}}]{Shrestha2022}%
  \BibitemOpen
  \bibfield  {author} {\bibinfo {author} {\bibfnamefont {S.}~\bibnamefont
  {Shrestha}}, \bibinfo {author} {\bibfnamefont {M.}~\bibnamefont
  {Krautloher}}, \bibinfo {author} {\bibfnamefont {M.}~\bibnamefont {Zhu}},
  \bibinfo {author} {\bibfnamefont {J.}~\bibnamefont {Kim}}, \bibinfo {author}
  {\bibfnamefont {J.}~\bibnamefont {Hwang}}, \bibinfo {author} {\bibfnamefont
  {J.}~\bibnamefont {Kim}}, \bibinfo {author} {\bibfnamefont {J.-W.}\
  \bibnamefont {Kim}}, \bibinfo {author} {\bibfnamefont {B.}~\bibnamefont
  {Keimer}},\ and\ \bibinfo {author} {\bibfnamefont {A.}~\bibnamefont {Seo}},\
  }\bibfield  {title} {\bibinfo {title} {{Emergent interlayer magnetic order
  via strain-induced orthorhombic distortion in the $5d$ Mott insulator
  ${\mathrm{Sr}}_{2}{\mathrm{IrO}}_{4}$}},\ }\href
  {https://doi.org/10.1103/PhysRevB.105.L100404} {\bibfield  {journal}
  {\bibinfo  {journal} {Phys. Rev. B}\ }\textbf {\bibinfo {volume} {105}},\
  \bibinfo {pages} {L100404} (\bibinfo {year} {2022})}\BibitemShut {NoStop}%
\bibitem [{\citenamefont {Engstr\"om}\ \emph {et~al.}(2023)\citenamefont
  {Engstr\"om}, \citenamefont {Liu}, \citenamefont {Witczak-Krempa},\ and\
  \citenamefont {Pereg-Barnea}}]{Engstrom2023}%
  \BibitemOpen
  \bibfield  {author} {\bibinfo {author} {\bibfnamefont {L.}~\bibnamefont
  {Engstr\"om}}, \bibinfo {author} {\bibfnamefont {C.-C.}\ \bibnamefont {Liu}},
  \bibinfo {author} {\bibfnamefont {W.}~\bibnamefont {Witczak-Krempa}},\ and\
  \bibinfo {author} {\bibfnamefont {T.}~\bibnamefont {Pereg-Barnea}},\
  }\bibfield  {title} {\bibinfo {title} {{Strain-induced superconductivity in
  ${\mathrm{Sr}}_{2}{\mathrm{IrO}}_{4}$}},\ }\href
  {https://doi.org/10.1103/PhysRevB.108.014508} {\bibfield  {journal} {\bibinfo
   {journal} {Phys. Rev. B}\ }\textbf {\bibinfo {volume} {108}},\ \bibinfo
  {pages} {014508} (\bibinfo {year} {2023})}\BibitemShut {NoStop}%
\bibitem [{\citenamefont {Moon}\ \emph {et~al.}(2008)\citenamefont {Moon},
  \citenamefont {Jin}, \citenamefont {Kim}, \citenamefont {Choi}, \citenamefont
  {Lee}, \citenamefont {Yu}, \citenamefont {Cao}, \citenamefont {Sumi},
  \citenamefont {Funakubo}, \citenamefont {Bernhard},\ and\ \citenamefont
  {Noh}}]{Moon2008}%
  \BibitemOpen
  \bibfield  {author} {\bibinfo {author} {\bibfnamefont {S.~J.}\ \bibnamefont
  {Moon}}, \bibinfo {author} {\bibfnamefont {H.}~\bibnamefont {Jin}}, \bibinfo
  {author} {\bibfnamefont {K.~W.}\ \bibnamefont {Kim}}, \bibinfo {author}
  {\bibfnamefont {W.~S.}\ \bibnamefont {Choi}}, \bibinfo {author}
  {\bibfnamefont {Y.~S.}\ \bibnamefont {Lee}}, \bibinfo {author} {\bibfnamefont
  {J.}~\bibnamefont {Yu}}, \bibinfo {author} {\bibfnamefont {G.}~\bibnamefont
  {Cao}}, \bibinfo {author} {\bibfnamefont {A.}~\bibnamefont {Sumi}}, \bibinfo
  {author} {\bibfnamefont {H.}~\bibnamefont {Funakubo}}, \bibinfo {author}
  {\bibfnamefont {C.}~\bibnamefont {Bernhard}},\ and\ \bibinfo {author}
  {\bibfnamefont {T.~W.}\ \bibnamefont {Noh}},\ }\bibfield  {title} {\bibinfo
  {title} {{Dimensionality-Controlled Insulator-Metal Transition and Correlated
  Metallic State in $5d$ Transition Metal Oxides
  ${\mathrm{Sr}}_{n+1}{\mathrm{Ir}}_{n}{\mathrm{O}}_{3n+1}$ ($n=1$, 2, and
  $\ensuremath{\infty}$)}},\ }\href
  {https://doi.org/10.1103/PhysRevLett.101.226402} {\bibfield  {journal}
  {\bibinfo  {journal} {Phys. Rev. Lett.}\ }\textbf {\bibinfo {volume} {101}},\
  \bibinfo {pages} {226402} (\bibinfo {year} {2008})}\BibitemShut {NoStop}%
\bibitem [{\citenamefont {Wang}\ \emph {et~al.}(2013)\citenamefont {Wang},
  \citenamefont {Cao}, \citenamefont {Waugh}, \citenamefont {Park},
  \citenamefont {Qi}, \citenamefont {Korneta}, \citenamefont {Cao},\ and\
  \citenamefont {Dessau}}]{Wang2013}%
  \BibitemOpen
  \bibfield  {author} {\bibinfo {author} {\bibfnamefont {Q.}~\bibnamefont
  {Wang}}, \bibinfo {author} {\bibfnamefont {Y.}~\bibnamefont {Cao}}, \bibinfo
  {author} {\bibfnamefont {J.~A.}\ \bibnamefont {Waugh}}, \bibinfo {author}
  {\bibfnamefont {S.~R.}\ \bibnamefont {Park}}, \bibinfo {author}
  {\bibfnamefont {T.~F.}\ \bibnamefont {Qi}}, \bibinfo {author} {\bibfnamefont
  {O.~B.}\ \bibnamefont {Korneta}}, \bibinfo {author} {\bibfnamefont
  {G.}~\bibnamefont {Cao}},\ and\ \bibinfo {author} {\bibfnamefont {D.~S.}\
  \bibnamefont {Dessau}},\ }\bibfield  {title} {\bibinfo {title}
  {{Dimensionality-controlled Mott transition and correlation effects in
  single-layer and bilayer perovskite iridates}},\ }\href
  {https://doi.org/10.1103/PhysRevB.87.245109} {\bibfield  {journal} {\bibinfo
  {journal} {Phys. Rev. B}\ }\textbf {\bibinfo {volume} {87}},\ \bibinfo
  {pages} {245109} (\bibinfo {year} {2013})}\BibitemShut {NoStop}%
\bibitem [{\citenamefont {Kim}\ \emph {et~al.}(2014{\natexlab{a}})\citenamefont
  {Kim}, \citenamefont {Krupin}, \citenamefont {Denlinger}, \citenamefont
  {Bostwick}, \citenamefont {Rotenberg}, \citenamefont {Zhao}, \citenamefont
  {Mitchell}, \citenamefont {Allen},\ and\ \citenamefont {Kim}}]{Kim2014SC}%
  \BibitemOpen
  \bibfield  {author} {\bibinfo {author} {\bibfnamefont {Y.~K.}\ \bibnamefont
  {Kim}}, \bibinfo {author} {\bibfnamefont {O.}~\bibnamefont {Krupin}},
  \bibinfo {author} {\bibfnamefont {J.}~\bibnamefont {Denlinger}}, \bibinfo
  {author} {\bibfnamefont {A.}~\bibnamefont {Bostwick}}, \bibinfo {author}
  {\bibfnamefont {E.}~\bibnamefont {Rotenberg}}, \bibinfo {author}
  {\bibfnamefont {Q.}~\bibnamefont {Zhao}}, \bibinfo {author} {\bibfnamefont
  {J.}~\bibnamefont {Mitchell}}, \bibinfo {author} {\bibfnamefont
  {J.}~\bibnamefont {Allen}},\ and\ \bibinfo {author} {\bibfnamefont
  {B.}~\bibnamefont {Kim}},\ }\bibfield  {title} {\bibinfo {title} {{Fermi arcs
  in a doped pseudospin-1/2 Heisenberg antiferromagnet}},\ }\href
  {https://doi.org/10.1126/science.1251} {\bibfield  {journal} {\bibinfo
  {journal} {Science}\ }\textbf {\bibinfo {volume} {345}},\ \bibinfo {pages}
  {187} (\bibinfo {year} {2014}{\natexlab{a}})}\BibitemShut {NoStop}%
\bibitem [{\citenamefont {Kim}\ \emph {et~al.}(2014{\natexlab{b}})\citenamefont
  {Kim}, \citenamefont {Daghofer}, \citenamefont {Said}, \citenamefont {Gog},
  \citenamefont {Van~den Brink}, \citenamefont {Khaliullin},\ and\
  \citenamefont {Kim}}]{Kim2014NC}%
  \BibitemOpen
  \bibfield  {author} {\bibinfo {author} {\bibfnamefont {J.}~\bibnamefont
  {Kim}}, \bibinfo {author} {\bibfnamefont {M.}~\bibnamefont {Daghofer}},
  \bibinfo {author} {\bibfnamefont {A.}~\bibnamefont {Said}}, \bibinfo {author}
  {\bibfnamefont {T.}~\bibnamefont {Gog}}, \bibinfo {author} {\bibfnamefont
  {J.}~\bibnamefont {Van~den Brink}}, \bibinfo {author} {\bibfnamefont
  {G.}~\bibnamefont {Khaliullin}},\ and\ \bibinfo {author} {\bibfnamefont
  {B.}~\bibnamefont {Kim}},\ }\bibfield  {title} {\bibinfo {title} {{Excitonic
  quasiparticles in a spin-orbit Mott insulator}},\ }\href
  {https://doi.org/10.1038/ncomms5453} {\bibfield  {journal} {\bibinfo
  {journal} {Nature communications}\ }\textbf {\bibinfo {volume} {5}},\
  \bibinfo {pages} {4453} (\bibinfo {year} {2014}{\natexlab{b}})}\BibitemShut
  {NoStop}%
\bibitem [{\citenamefont {Zhao}\ \emph {et~al.}(2016)\citenamefont {Zhao},
  \citenamefont {Torchinsky}, \citenamefont {Chu}, \citenamefont {Ivanov},
  \citenamefont {Lifshitz}, \citenamefont {Flint}, \citenamefont {Qi},
  \citenamefont {Cao},\ and\ \citenamefont {Hsieh}}]{Zhao2016}%
  \BibitemOpen
  \bibfield  {author} {\bibinfo {author} {\bibfnamefont {L.}~\bibnamefont
  {Zhao}}, \bibinfo {author} {\bibfnamefont {D.}~\bibnamefont {Torchinsky}},
  \bibinfo {author} {\bibfnamefont {H.}~\bibnamefont {Chu}}, \bibinfo {author}
  {\bibfnamefont {V.}~\bibnamefont {Ivanov}}, \bibinfo {author} {\bibfnamefont
  {R.}~\bibnamefont {Lifshitz}}, \bibinfo {author} {\bibfnamefont
  {R.}~\bibnamefont {Flint}}, \bibinfo {author} {\bibfnamefont
  {T.}~\bibnamefont {Qi}}, \bibinfo {author} {\bibfnamefont {G.}~\bibnamefont
  {Cao}},\ and\ \bibinfo {author} {\bibfnamefont {D.}~\bibnamefont {Hsieh}},\
  }\bibfield  {title} {\bibinfo {title} {Evidence of an odd-parity hidden order
  in a spin-orbit coupled correlated iridate},\ }\href
  {https://doi.org/10.1038/nphys3517} {\bibfield  {journal} {\bibinfo
  {journal} {Nature Physics}\ }\textbf {\bibinfo {volume} {12}},\ \bibinfo
  {pages} {32} (\bibinfo {year} {2016})}\BibitemShut {NoStop}%
\bibitem [{\citenamefont {Zhou}\ \emph {et~al.}(2017)\citenamefont {Zhou},
  \citenamefont {Jiang}, \citenamefont {Chen},\ and\ \citenamefont
  {Wang}}]{Zhou2017}%
  \BibitemOpen
  \bibfield  {author} {\bibinfo {author} {\bibfnamefont {S.}~\bibnamefont
  {Zhou}}, \bibinfo {author} {\bibfnamefont {K.}~\bibnamefont {Jiang}},
  \bibinfo {author} {\bibfnamefont {H.}~\bibnamefont {Chen}},\ and\ \bibinfo
  {author} {\bibfnamefont {Z.}~\bibnamefont {Wang}},\ }\bibfield  {title}
  {\bibinfo {title} {{Correlation Effects and Hidden Spin-Orbit Entangled
  Electronic Order in Parent and Electron-Doped Iridates
  ${\mathrm{Sr}}_{2}{\mathrm{IrO}}_{4}$}},\ }\href
  {https://doi.org/10.1103/PhysRevX.7.041018} {\bibfield  {journal} {\bibinfo
  {journal} {Phys. Rev. X}\ }\textbf {\bibinfo {volume} {7}},\ \bibinfo {pages}
  {041018} (\bibinfo {year} {2017})}\BibitemShut {NoStop}%
\bibitem [{\citenamefont {Jeong}\ \emph {et~al.}(2017)\citenamefont {Jeong},
  \citenamefont {Sidis}, \citenamefont {Louat}, \citenamefont {Brouet},\ and\
  \citenamefont {Bourges}}]{Jeong2017}%
  \BibitemOpen
  \bibfield  {author} {\bibinfo {author} {\bibfnamefont {J.}~\bibnamefont
  {Jeong}}, \bibinfo {author} {\bibfnamefont {Y.}~\bibnamefont {Sidis}},
  \bibinfo {author} {\bibfnamefont {A.}~\bibnamefont {Louat}}, \bibinfo
  {author} {\bibfnamefont {V.}~\bibnamefont {Brouet}},\ and\ \bibinfo {author}
  {\bibfnamefont {P.}~\bibnamefont {Bourges}},\ }\bibfield  {title} {\bibinfo
  {title} {{Time-reversal symmetry breaking hidden order in
  Sr$_2$(Ir,Rh)O$_4$}},\ }\href {https://doi.org/10.1038/ncomms15119}
  {\bibfield  {journal} {\bibinfo  {journal} {Nat. Commun.}\ }\textbf {\bibinfo
  {volume} {8}},\ \bibinfo {pages} {15119} (\bibinfo {year} {2017})},\ \Eprint
  {https://arxiv.org/abs/1701.06485} {1701.06485} \BibitemShut {NoStop}%
\bibitem [{\citenamefont {Wang}\ \emph {et~al.}(2018)\citenamefont {Wang},
  \citenamefont {Bachar}, \citenamefont {Teyssier}, \citenamefont {Luo},
  \citenamefont {Rischau}, \citenamefont {Scheerer}, \citenamefont {de~la
  Torre}, \citenamefont {Perry}, \citenamefont {Baumberger},\ and\
  \citenamefont {van~der Marel}}]{Wang2018}%
  \BibitemOpen
  \bibfield  {author} {\bibinfo {author} {\bibfnamefont {K.}~\bibnamefont
  {Wang}}, \bibinfo {author} {\bibfnamefont {N.}~\bibnamefont {Bachar}},
  \bibinfo {author} {\bibfnamefont {J.}~\bibnamefont {Teyssier}}, \bibinfo
  {author} {\bibfnamefont {W.}~\bibnamefont {Luo}}, \bibinfo {author}
  {\bibfnamefont {C.~W.}\ \bibnamefont {Rischau}}, \bibinfo {author}
  {\bibfnamefont {G.}~\bibnamefont {Scheerer}}, \bibinfo {author}
  {\bibfnamefont {A.}~\bibnamefont {de~la Torre}}, \bibinfo {author}
  {\bibfnamefont {R.~S.}\ \bibnamefont {Perry}}, \bibinfo {author}
  {\bibfnamefont {F.}~\bibnamefont {Baumberger}},\ and\ \bibinfo {author}
  {\bibfnamefont {D.}~\bibnamefont {van~der Marel}},\ }\bibfield  {title}
  {\bibinfo {title} {{Mott transition and collective charge pinning in electron
  doped ${\mathrm{Sr}}_{2}{\mathrm{IrO}}_{4}$}},\ }\href
  {https://doi.org/10.1103/PhysRevB.98.045107} {\bibfield  {journal} {\bibinfo
  {journal} {Phys. Rev. B}\ }\textbf {\bibinfo {volume} {98}},\ \bibinfo
  {pages} {045107} (\bibinfo {year} {2018})}\BibitemShut {NoStop}%
\bibitem [{\citenamefont {Cao}\ \emph {et~al.}(2018)\citenamefont {Cao},
  \citenamefont {Terzic}, \citenamefont {Zhao}, \citenamefont {Zheng},
  \citenamefont {De~Long},\ and\ \citenamefont {Riseborough}}]{Cao2018}%
  \BibitemOpen
  \bibfield  {author} {\bibinfo {author} {\bibfnamefont {G.}~\bibnamefont
  {Cao}}, \bibinfo {author} {\bibfnamefont {J.}~\bibnamefont {Terzic}},
  \bibinfo {author} {\bibfnamefont {H.~D.}\ \bibnamefont {Zhao}}, \bibinfo
  {author} {\bibfnamefont {H.}~\bibnamefont {Zheng}}, \bibinfo {author}
  {\bibfnamefont {L.~E.}\ \bibnamefont {De~Long}},\ and\ \bibinfo {author}
  {\bibfnamefont {P.~S.}\ \bibnamefont {Riseborough}},\ }\bibfield  {title}
  {\bibinfo {title} {{Electrical Control of Structural and Physical Properties
  via Strong Spin-Orbit Interactions in
  ${\mathrm{Sr}}_{2}{\mathrm{IrO}}_{4}$}},\ }\href
  {https://doi.org/10.1103/PhysRevLett.120.017201} {\bibfield  {journal}
  {\bibinfo  {journal} {Phys. Rev. Lett.}\ }\textbf {\bibinfo {volume} {120}},\
  \bibinfo {pages} {017201} (\bibinfo {year} {2018})}\BibitemShut {NoStop}%
\bibitem [{\citenamefont {Murayama}\ \emph {et~al.}(2021)\citenamefont
  {Murayama}, \citenamefont {Ishida}, \citenamefont {Kurihara}, \citenamefont
  {Ono}, \citenamefont {Sato}, \citenamefont {Kasahara}, \citenamefont
  {Watanabe}, \citenamefont {Yanase}, \citenamefont {Cao}, \citenamefont
  {Mizukami}, \citenamefont {Shibauchi}, \citenamefont {Matsuda},\ and\
  \citenamefont {Kasahara}}]{Murayama2021}%
  \BibitemOpen
  \bibfield  {author} {\bibinfo {author} {\bibfnamefont {H.}~\bibnamefont
  {Murayama}}, \bibinfo {author} {\bibfnamefont {K.}~\bibnamefont {Ishida}},
  \bibinfo {author} {\bibfnamefont {R.}~\bibnamefont {Kurihara}}, \bibinfo
  {author} {\bibfnamefont {T.}~\bibnamefont {Ono}}, \bibinfo {author}
  {\bibfnamefont {Y.}~\bibnamefont {Sato}}, \bibinfo {author} {\bibfnamefont
  {Y.}~\bibnamefont {Kasahara}}, \bibinfo {author} {\bibfnamefont
  {H.}~\bibnamefont {Watanabe}}, \bibinfo {author} {\bibfnamefont
  {Y.}~\bibnamefont {Yanase}}, \bibinfo {author} {\bibfnamefont
  {G.}~\bibnamefont {Cao}}, \bibinfo {author} {\bibfnamefont {Y.}~\bibnamefont
  {Mizukami}}, \bibinfo {author} {\bibfnamefont {T.}~\bibnamefont {Shibauchi}},
  \bibinfo {author} {\bibfnamefont {Y.}~\bibnamefont {Matsuda}},\ and\ \bibinfo
  {author} {\bibfnamefont {S.}~\bibnamefont {Kasahara}},\ }\bibfield  {title}
  {\bibinfo {title} {{Bond Directional Anapole Order in a Spin-Orbit Coupled
  Mott Insulator
  ${\mathrm{Sr}}_{2}({\mathrm{Ir}}_{1\ensuremath{-}x}{\mathrm{Rh}}_{x}){\mathrm{O}}_{4}$}},\
  }\href {https://doi.org/10.1103/PhysRevX.11.011021} {\bibfield  {journal}
  {\bibinfo  {journal} {Phys. Rev. X}\ }\textbf {\bibinfo {volume} {11}},\
  \bibinfo {pages} {011021} (\bibinfo {year} {2021})}\BibitemShut {NoStop}%
\bibitem [{\citenamefont {Kim}\ \emph {et~al.}(2024{\natexlab{a}})\citenamefont
  {Kim}, \citenamefont {Kim}, \citenamefont {Kwon}, \citenamefont {Kim},
  \citenamefont {Kim}, \citenamefont {Ha}, \citenamefont {Kim}, \citenamefont
  {Lee}, \citenamefont {Kim}, \citenamefont {Cho}, \citenamefont {Heo},
  \citenamefont {Jang}, \citenamefont {Sahle}, \citenamefont {Longo},
  \citenamefont {Strempfer}, \citenamefont {Fabbris}, \citenamefont {Choi},
  \citenamefont {Haskel}, \citenamefont {Kim}, \citenamefont {Kim},\ and\
  \citenamefont {Kim}}]{Kim2024}%
  \BibitemOpen
  \bibfield  {author} {\bibinfo {author} {\bibfnamefont {H.}~\bibnamefont
  {Kim}}, \bibinfo {author} {\bibfnamefont {J.~K.}\ \bibnamefont {Kim}},
  \bibinfo {author} {\bibfnamefont {J.}~\bibnamefont {Kwon}}, \bibinfo {author}
  {\bibfnamefont {J.}~\bibnamefont {Kim}}, \bibinfo {author} {\bibfnamefont
  {H.~W.~J.}\ \bibnamefont {Kim}}, \bibinfo {author} {\bibfnamefont
  {S.}~\bibnamefont {Ha}}, \bibinfo {author} {\bibfnamefont {K.}~\bibnamefont
  {Kim}}, \bibinfo {author} {\bibfnamefont {W.}~\bibnamefont {Lee}}, \bibinfo
  {author} {\bibfnamefont {J.}~\bibnamefont {Kim}}, \bibinfo {author}
  {\bibfnamefont {G.~Y.}\ \bibnamefont {Cho}}, \bibinfo {author} {\bibfnamefont
  {H.}~\bibnamefont {Heo}}, \bibinfo {author} {\bibfnamefont {J.}~\bibnamefont
  {Jang}}, \bibinfo {author} {\bibfnamefont {C.~J.}\ \bibnamefont {Sahle}},
  \bibinfo {author} {\bibfnamefont {A.}~\bibnamefont {Longo}}, \bibinfo
  {author} {\bibfnamefont {J.}~\bibnamefont {Strempfer}}, \bibinfo {author}
  {\bibfnamefont {G.}~\bibnamefont {Fabbris}}, \bibinfo {author} {\bibfnamefont
  {Y.}~\bibnamefont {Choi}}, \bibinfo {author} {\bibfnamefont {D.}~\bibnamefont
  {Haskel}}, \bibinfo {author} {\bibfnamefont {J.}~\bibnamefont {Kim}},
  \bibinfo {author} {\bibfnamefont {J.~W.}\ \bibnamefont {Kim}},\ and\ \bibinfo
  {author} {\bibfnamefont {B.~J.}\ \bibnamefont {Kim}},\ }\bibfield  {title}
  {\bibinfo {title} {{Quantum spin nematic phase in a square-lattice
  iridate}},\ }\href {https://doi.org/10.1038/s41586-023-06829-4} {\bibfield
  {journal} {\bibinfo  {journal} {Nature}\ }\textbf {\bibinfo {volume} {625}},\
  \bibinfo {pages} {264} (\bibinfo {year} {2024}{\natexlab{a}})}\BibitemShut
  {NoStop}%
\bibitem [{\citenamefont {Battisti}\ \emph {et~al.}(2017)\citenamefont
  {Battisti}, \citenamefont {Bastiaans}, \citenamefont {Fedoseev},
  \citenamefont {de~la Torre}, \citenamefont {Iliopoulos}, \citenamefont
  {Tamai}, \citenamefont {Hunter}, \citenamefont {Perry}, \citenamefont
  {Zaanen}, \citenamefont {Baumberger},\ and\ \citenamefont
  {Allan}}]{Battisti2017}%
  \BibitemOpen
  \bibfield  {author} {\bibinfo {author} {\bibfnamefont {I.}~\bibnamefont
  {Battisti}}, \bibinfo {author} {\bibfnamefont {K.~M.}\ \bibnamefont
  {Bastiaans}}, \bibinfo {author} {\bibfnamefont {V.}~\bibnamefont {Fedoseev}},
  \bibinfo {author} {\bibfnamefont {A.}~\bibnamefont {de~la Torre}}, \bibinfo
  {author} {\bibfnamefont {N.}~\bibnamefont {Iliopoulos}}, \bibinfo {author}
  {\bibfnamefont {A.}~\bibnamefont {Tamai}}, \bibinfo {author} {\bibfnamefont
  {E.~C.}\ \bibnamefont {Hunter}}, \bibinfo {author} {\bibfnamefont {R.~S.}\
  \bibnamefont {Perry}}, \bibinfo {author} {\bibfnamefont {J.}~\bibnamefont
  {Zaanen}}, \bibinfo {author} {\bibfnamefont {F.}~\bibnamefont {Baumberger}},\
  and\ \bibinfo {author} {\bibfnamefont {M.~P.}\ \bibnamefont {Allan}},\
  }\bibfield  {title} {\bibinfo {title} {{Universality of pseudogap and
  emergent order in lightly doped Mott insulators}},\ }\href
  {https://doi.org/10.1038/nphys3894} {\bibfield  {journal} {\bibinfo
  {journal} {Nature Physics}\ }\textbf {\bibinfo {volume} {13}},\ \bibinfo
  {pages} {21} (\bibinfo {year} {2017})}\BibitemShut {NoStop}%
\bibitem [{\citenamefont {Seo}\ \emph {et~al.}(2017)\citenamefont {Seo},
  \citenamefont {Ahn}, \citenamefont {Song}, \citenamefont {Chen},
  \citenamefont {Wilson},\ and\ \citenamefont {Moon}}]{Seo2017}%
  \BibitemOpen
  \bibfield  {author} {\bibinfo {author} {\bibfnamefont {J.~H.}\ \bibnamefont
  {Seo}}, \bibinfo {author} {\bibfnamefont {G.~H.}\ \bibnamefont {Ahn}},
  \bibinfo {author} {\bibfnamefont {S.~J.}\ \bibnamefont {Song}}, \bibinfo
  {author} {\bibfnamefont {X.}~\bibnamefont {Chen}}, \bibinfo {author}
  {\bibfnamefont {S.~D.}\ \bibnamefont {Wilson}},\ and\ \bibinfo {author}
  {\bibfnamefont {S.~J.}\ \bibnamefont {Moon}},\ }\bibfield  {title} {\bibinfo
  {title} {{Infrared probe of pseudogap in electron-doped Sr$_2$IrO$_4$}},\
  }\href {https://doi.org/10.1038/s41598-017-10725-z} {\bibfield  {journal}
  {\bibinfo  {journal} {Sci. Rep.}\ }\textbf {\bibinfo {volume} {7}},\ \bibinfo
  {pages} {10494} (\bibinfo {year} {2017})}\BibitemShut {NoStop}%
\bibitem [{\citenamefont {Louat}\ \emph {et~al.}(2019)\citenamefont {Louat},
  \citenamefont {Lenz}, \citenamefont {Biermann}, \citenamefont {Martins},
  \citenamefont {Bertran}, \citenamefont {Le~F\`evre}, \citenamefont {Rault},
  \citenamefont {Bert},\ and\ \citenamefont {Brouet}}]{Louat2019}%
  \BibitemOpen
  \bibfield  {author} {\bibinfo {author} {\bibfnamefont {A.}~\bibnamefont
  {Louat}}, \bibinfo {author} {\bibfnamefont {B.}~\bibnamefont {Lenz}},
  \bibinfo {author} {\bibfnamefont {S.}~\bibnamefont {Biermann}}, \bibinfo
  {author} {\bibfnamefont {C.}~\bibnamefont {Martins}}, \bibinfo {author}
  {\bibfnamefont {F.~m.~c.}\ \bibnamefont {Bertran}}, \bibinfo {author}
  {\bibfnamefont {P.}~\bibnamefont {Le~F\`evre}}, \bibinfo {author}
  {\bibfnamefont {J.~E.}\ \bibnamefont {Rault}}, \bibinfo {author}
  {\bibfnamefont {F.}~\bibnamefont {Bert}},\ and\ \bibinfo {author}
  {\bibfnamefont {V.}~\bibnamefont {Brouet}},\ }\bibfield  {title} {\bibinfo
  {title} {{ARPES study of orbital character, symmetry breaking, and pseudogaps
  in doped and pure ${\mathrm{Sr}}_{2}{\mathrm{IrO}}_{4}$}},\ }\href
  {https://doi.org/10.1103/PhysRevB.100.205135} {\bibfield  {journal} {\bibinfo
   {journal} {Phys. Rev. B}\ }\textbf {\bibinfo {volume} {100}},\ \bibinfo
  {pages} {205135} (\bibinfo {year} {2019})}\BibitemShut {NoStop}%
\bibitem [{\citenamefont {Wang}\ and\ \citenamefont
  {Senthil}(2011)}]{Wang2011}%
  \BibitemOpen
  \bibfield  {author} {\bibinfo {author} {\bibfnamefont {F.}~\bibnamefont
  {Wang}}\ and\ \bibinfo {author} {\bibfnamefont {T.}~\bibnamefont {Senthil}},\
  }\bibfield  {title} {\bibinfo {title} {{Twisted Hubbard Model for
  ${\mathrm{Sr}}_{2}{\mathrm{IrO}}_{4}$: Magnetism and Possible High
  Temperature Superconductivity}},\ }\href
  {https://doi.org/10.1103/PhysRevLett.106.136402} {\bibfield  {journal}
  {\bibinfo  {journal} {Phys. Rev. Lett.}\ }\textbf {\bibinfo {volume} {106}},\
  \bibinfo {pages} {136402} (\bibinfo {year} {2011})}\BibitemShut {NoStop}%
\bibitem [{\citenamefont {Yan}\ \emph {et~al.}(2015)\citenamefont {Yan},
  \citenamefont {Ren}, \citenamefont {Xu}, \citenamefont {Xie}, \citenamefont
  {Tao}, \citenamefont {Choi}, \citenamefont {Lee}, \citenamefont {Choi},
  \citenamefont {Zhang},\ and\ \citenamefont {Feng}}]{Yan2015}%
  \BibitemOpen
  \bibfield  {author} {\bibinfo {author} {\bibfnamefont {Y.~J.}\ \bibnamefont
  {Yan}}, \bibinfo {author} {\bibfnamefont {M.~Q.}\ \bibnamefont {Ren}},
  \bibinfo {author} {\bibfnamefont {H.~C.}\ \bibnamefont {Xu}}, \bibinfo
  {author} {\bibfnamefont {B.~P.}\ \bibnamefont {Xie}}, \bibinfo {author}
  {\bibfnamefont {R.}~\bibnamefont {Tao}}, \bibinfo {author} {\bibfnamefont
  {H.~Y.}\ \bibnamefont {Choi}}, \bibinfo {author} {\bibfnamefont
  {N.}~\bibnamefont {Lee}}, \bibinfo {author} {\bibfnamefont {Y.~J.}\
  \bibnamefont {Choi}}, \bibinfo {author} {\bibfnamefont {T.}~\bibnamefont
  {Zhang}},\ and\ \bibinfo {author} {\bibfnamefont {D.~L.}\ \bibnamefont
  {Feng}},\ }\bibfield  {title} {\bibinfo {title} {{Electron-Doped
  ${\mathrm{Sr}}_{2}{\mathrm{IrO}}_{4}$: An Analogue of Hole-Doped Cuprate
  Superconductors Demonstrated by Scanning Tunneling Microscopy}},\ }\href
  {https://doi.org/10.1103/PhysRevX.5.041018} {\bibfield  {journal} {\bibinfo
  {journal} {Phys. Rev. X}\ }\textbf {\bibinfo {volume} {5}},\ \bibinfo {pages}
  {041018} (\bibinfo {year} {2015})}\BibitemShut {NoStop}%
\bibitem [{\citenamefont {de'Medici}\ \emph {et~al.}(2005)\citenamefont
  {de'Medici}, \citenamefont {Georges},\ and\ \citenamefont
  {Biermann}}]{Medici2005}%
  \BibitemOpen
  \bibfield  {author} {\bibinfo {author} {\bibfnamefont {L.}~\bibnamefont
  {de'Medici}}, \bibinfo {author} {\bibfnamefont {A.}~\bibnamefont {Georges}},\
  and\ \bibinfo {author} {\bibfnamefont {S.}~\bibnamefont {Biermann}},\
  }\bibfield  {title} {\bibinfo {title} {{Orbital-selective Mott transition in
  multiband systems: Slave-spin representation and dynamical mean-field
  theory}},\ }\href {https://doi.org/10.1103/PhysRevB.72.205124} {\bibfield
  {journal} {\bibinfo  {journal} {Phys. Rev. B}\ }\textbf {\bibinfo {volume}
  {72}},\ \bibinfo {pages} {205124} (\bibinfo {year} {2005})}\BibitemShut
  {NoStop}%
\bibitem [{\citenamefont {de' Medici}\ \emph {et~al.}(2009)\citenamefont {de'
  Medici}, \citenamefont {Hassan}, \citenamefont {Capone},\ and\ \citenamefont
  {Dai}}]{Medici2009}%
  \BibitemOpen
  \bibfield  {author} {\bibinfo {author} {\bibfnamefont {L.}~\bibnamefont {de'
  Medici}}, \bibinfo {author} {\bibfnamefont {S.~R.}\ \bibnamefont {Hassan}},
  \bibinfo {author} {\bibfnamefont {M.}~\bibnamefont {Capone}},\ and\ \bibinfo
  {author} {\bibfnamefont {X.}~\bibnamefont {Dai}},\ }\bibfield  {title}
  {\bibinfo {title} {{Orbital-Selective Mott Transition out of Band Degeneracy
  Lifting}},\ }\href {https://doi.org/10.1103/PhysRevLett.102.126401}
  {\bibfield  {journal} {\bibinfo  {journal} {Phys. Rev. Lett.}\ }\textbf
  {\bibinfo {volume} {102}},\ \bibinfo {pages} {126401} (\bibinfo {year}
  {2009})}\BibitemShut {NoStop}%
\bibitem [{\citenamefont {de' Medici}\ \emph {et~al.}(2014)\citenamefont {de'
  Medici}, \citenamefont {Giovannetti},\ and\ \citenamefont
  {Capone}}]{Medici2014}%
  \BibitemOpen
  \bibfield  {author} {\bibinfo {author} {\bibfnamefont {L.}~\bibnamefont {de'
  Medici}}, \bibinfo {author} {\bibfnamefont {G.}~\bibnamefont {Giovannetti}},\
  and\ \bibinfo {author} {\bibfnamefont {M.}~\bibnamefont {Capone}},\
  }\bibfield  {title} {\bibinfo {title} {Selective mott physics as a key to
  iron superconductors},\ }\href
  {https://doi.org/10.1103/PhysRevLett.112.177001} {\bibfield  {journal}
  {\bibinfo  {journal} {Phys. Rev. Lett.}\ }\textbf {\bibinfo {volume} {112}},\
  \bibinfo {pages} {177001} (\bibinfo {year} {2014})}\BibitemShut {NoStop}%
\bibitem [{\citenamefont {Yu}\ and\ \citenamefont {Si}(2013)}]{Yu2013PRL}%
  \BibitemOpen
  \bibfield  {author} {\bibinfo {author} {\bibfnamefont {R.}~\bibnamefont
  {Yu}}\ and\ \bibinfo {author} {\bibfnamefont {Q.}~\bibnamefont {Si}},\
  }\bibfield  {title} {\bibinfo {title} {{Orbital-Selective Mott Phase in
  Multiorbital Models for Alkaline Iron Selenides
  ${\mathrm{K}}_{1\ensuremath{-}x}{\mathrm{Fe}}_{2\ensuremath{-}y}{\mathrm{Se}}_{2}$}},\
  }\href {https://doi.org/10.1103/PhysRevLett.110.146402} {\bibfield  {journal}
  {\bibinfo  {journal} {Phys. Rev. Lett.}\ }\textbf {\bibinfo {volume} {110}},\
  \bibinfo {pages} {146402} (\bibinfo {year} {2013})}\BibitemShut {NoStop}%
\bibitem [{\citenamefont {Yu}\ and\ \citenamefont {Si}(2017)}]{Yu2017PRB}%
  \BibitemOpen
  \bibfield  {author} {\bibinfo {author} {\bibfnamefont {R.}~\bibnamefont
  {Yu}}\ and\ \bibinfo {author} {\bibfnamefont {Q.}~\bibnamefont {Si}},\
  }\bibfield  {title} {\bibinfo {title} {Orbital-selective mott phase in
  multiorbital models for iron pnictides and chalcogenides},\ }\href
  {https://doi.org/10.1103/PhysRevB.96.125110} {\bibfield  {journal} {\bibinfo
  {journal} {Phys. Rev. B}\ }\textbf {\bibinfo {volume} {96}},\ \bibinfo
  {pages} {125110} (\bibinfo {year} {2017})}\BibitemShut {NoStop}%
\bibitem [{\citenamefont {Kim}\ \emph {et~al.}(2024{\natexlab{b}})\citenamefont
  {Kim}, \citenamefont {Choi}, \citenamefont {Brito},\ and\ \citenamefont
  {Kotliar}}]{Kim2024PRL}%
  \BibitemOpen
  \bibfield  {author} {\bibinfo {author} {\bibfnamefont {M.}~\bibnamefont
  {Kim}}, \bibinfo {author} {\bibfnamefont {S.}~\bibnamefont {Choi}}, \bibinfo
  {author} {\bibfnamefont {W.~H.}\ \bibnamefont {Brito}},\ and\ \bibinfo
  {author} {\bibfnamefont {G.}~\bibnamefont {Kotliar}},\ }\bibfield  {title}
  {\bibinfo {title} {{Orbital-Selective Mott Transition Effects and Nontrivial
  Topology of Iron Chalcogenide}},\ }\href
  {https://doi.org/10.1103/PhysRevLett.132.136504} {\bibfield  {journal}
  {\bibinfo  {journal} {Phys. Rev. Lett.}\ }\textbf {\bibinfo {volume} {132}},\
  \bibinfo {pages} {136504} (\bibinfo {year} {2024}{\natexlab{b}})}\BibitemShut
  {NoStop}%
\bibitem [{\citenamefont {Wang}\ \emph {et~al.}(2004)\citenamefont {Wang},
  \citenamefont {Yang}, \citenamefont {Sekharan}, \citenamefont {Souma},
  \citenamefont {Matsui}, \citenamefont {Sato}, \citenamefont {Takahashi},
  \citenamefont {Lu}, \citenamefont {Zhang}, \citenamefont {Jin}, \citenamefont
  {Mandrus}, \citenamefont {Plummer}, \citenamefont {Wang},\ and\ \citenamefont
  {Ding}}]{Wang2004PRL}%
  \BibitemOpen
  \bibfield  {author} {\bibinfo {author} {\bibfnamefont {S.-C.}\ \bibnamefont
  {Wang}}, \bibinfo {author} {\bibfnamefont {H.-B.}\ \bibnamefont {Yang}},
  \bibinfo {author} {\bibfnamefont {A.~K.~P.}\ \bibnamefont {Sekharan}},
  \bibinfo {author} {\bibfnamefont {S.}~\bibnamefont {Souma}}, \bibinfo
  {author} {\bibfnamefont {H.}~\bibnamefont {Matsui}}, \bibinfo {author}
  {\bibfnamefont {T.}~\bibnamefont {Sato}}, \bibinfo {author} {\bibfnamefont
  {T.}~\bibnamefont {Takahashi}}, \bibinfo {author} {\bibfnamefont
  {C.}~\bibnamefont {Lu}}, \bibinfo {author} {\bibfnamefont {J.}~\bibnamefont
  {Zhang}}, \bibinfo {author} {\bibfnamefont {R.}~\bibnamefont {Jin}}, \bibinfo
  {author} {\bibfnamefont {D.}~\bibnamefont {Mandrus}}, \bibinfo {author}
  {\bibfnamefont {E.~W.}\ \bibnamefont {Plummer}}, \bibinfo {author}
  {\bibfnamefont {Z.}~\bibnamefont {Wang}},\ and\ \bibinfo {author}
  {\bibfnamefont {H.}~\bibnamefont {Ding}},\ }\bibfield  {title} {\bibinfo
  {title} {{Fermi Surface Topology of
  ${\mathrm{C}\mathrm{a}}_{1.5}{\mathrm{S}\mathrm{r}}_{0.5}{\mathrm{R}\mathrm{u}\mathrm{O}}_{4}$
  Determined by Angle-Resolved Photoelectron Spectroscopy}},\ }\href
  {https://doi.org/10.1103/PhysRevLett.93.177007} {\bibfield  {journal}
  {\bibinfo  {journal} {Phys. Rev. Lett.}\ }\textbf {\bibinfo {volume} {93}},\
  \bibinfo {pages} {177007} (\bibinfo {year} {2004})}\BibitemShut {NoStop}%
\bibitem [{\citenamefont {Neupane}\ \emph {et~al.}(2009)\citenamefont
  {Neupane}, \citenamefont {Richard}, \citenamefont {Pan}, \citenamefont {Xu},
  \citenamefont {Jin}, \citenamefont {Mandrus}, \citenamefont {Dai},
  \citenamefont {Fang}, \citenamefont {Wang},\ and\ \citenamefont
  {Ding}}]{Neupane2009PRL}%
  \BibitemOpen
  \bibfield  {author} {\bibinfo {author} {\bibfnamefont {M.}~\bibnamefont
  {Neupane}}, \bibinfo {author} {\bibfnamefont {P.}~\bibnamefont {Richard}},
  \bibinfo {author} {\bibfnamefont {Z.-H.}\ \bibnamefont {Pan}}, \bibinfo
  {author} {\bibfnamefont {Y.-M.}\ \bibnamefont {Xu}}, \bibinfo {author}
  {\bibfnamefont {R.}~\bibnamefont {Jin}}, \bibinfo {author} {\bibfnamefont
  {D.}~\bibnamefont {Mandrus}}, \bibinfo {author} {\bibfnamefont
  {X.}~\bibnamefont {Dai}}, \bibinfo {author} {\bibfnamefont {Z.}~\bibnamefont
  {Fang}}, \bibinfo {author} {\bibfnamefont {Z.}~\bibnamefont {Wang}},\ and\
  \bibinfo {author} {\bibfnamefont {H.}~\bibnamefont {Ding}},\ }\bibfield
  {title} {\bibinfo {title} {{Observation of a Novel Orbital Selective Mott
  Transition in ${\mathrm{Ca}}_{1.8}{\mathrm{Sr}}_{0.2}{\mathrm{RuO}}_{4}$}},\
  }\href {https://doi.org/10.1103/PhysRevLett.103.097001} {\bibfield  {journal}
  {\bibinfo  {journal} {Phys. Rev. Lett.}\ }\textbf {\bibinfo {volume} {103}},\
  \bibinfo {pages} {097001} (\bibinfo {year} {2009})}\BibitemShut {NoStop}%
\bibitem [{Cao(2018)}]{Cao2018ROPP}%
  \BibitemOpen
  \bibfield  {title} {\bibinfo {title} {{The challenge of spin-orbit-tuned
  ground states in iridates: A key issues review}},\ }\href
  {https://doi.org/10.1088/1361-6633/aaa979} {\bibfield  {journal} {\bibinfo
  {journal} {Reports on Progress in Physics}\ }\textbf {\bibinfo {volume}
  {81}},\ \bibinfo {pages} {042502} (\bibinfo {year} {2018})}\BibitemShut
  {NoStop}%
\bibitem [{\citenamefont {Mazin}\ \emph {et~al.}(2012)\citenamefont {Mazin},
  \citenamefont {Jeschke}, \citenamefont {Foyevtsova}, \citenamefont
  {Valent\'{\i}},\ and\ \citenamefont {Khomskii}}]{Mazin2012PRL}%
  \BibitemOpen
  \bibfield  {author} {\bibinfo {author} {\bibfnamefont {I.~I.}\ \bibnamefont
  {Mazin}}, \bibinfo {author} {\bibfnamefont {H.~O.}\ \bibnamefont {Jeschke}},
  \bibinfo {author} {\bibfnamefont {K.}~\bibnamefont {Foyevtsova}}, \bibinfo
  {author} {\bibfnamefont {R.}~\bibnamefont {Valent\'{\i}}},\ and\ \bibinfo
  {author} {\bibfnamefont {D.~I.}\ \bibnamefont {Khomskii}},\ }\bibfield
  {title} {\bibinfo {title} {{${\mathrm{Na}}_{2}{\mathrm{IrO}}_{3}$ as a
  Molecular Orbital Crystal}},\ }\href
  {https://doi.org/10.1103/PhysRevLett.109.197201} {\bibfield  {journal}
  {\bibinfo  {journal} {Phys. Rev. Lett.}\ }\textbf {\bibinfo {volume} {109}},\
  \bibinfo {pages} {197201} (\bibinfo {year} {2012})}\BibitemShut {NoStop}%
\bibitem [{\citenamefont {Liu}\ \emph {et~al.}(2012)\citenamefont {Liu},
  \citenamefont {Katukuri}, \citenamefont {Hozoi}, \citenamefont {Yin},
  \citenamefont {Dean}, \citenamefont {Upton}, \citenamefont {Kim},
  \citenamefont {Casa}, \citenamefont {Said}, \citenamefont {Gog},
  \citenamefont {Qi}, \citenamefont {Cao}, \citenamefont {Tsvelik},
  \citenamefont {van~den Brink},\ and\ \citenamefont {Hill}}]{Liu2012PRL}%
  \BibitemOpen
  \bibfield  {author} {\bibinfo {author} {\bibfnamefont {X.}~\bibnamefont
  {Liu}}, \bibinfo {author} {\bibfnamefont {V.~M.}\ \bibnamefont {Katukuri}},
  \bibinfo {author} {\bibfnamefont {L.}~\bibnamefont {Hozoi}}, \bibinfo
  {author} {\bibfnamefont {W.-G.}\ \bibnamefont {Yin}}, \bibinfo {author}
  {\bibfnamefont {M.~P.~M.}\ \bibnamefont {Dean}}, \bibinfo {author}
  {\bibfnamefont {M.~H.}\ \bibnamefont {Upton}}, \bibinfo {author}
  {\bibfnamefont {J.}~\bibnamefont {Kim}}, \bibinfo {author} {\bibfnamefont
  {D.}~\bibnamefont {Casa}}, \bibinfo {author} {\bibfnamefont {A.}~\bibnamefont
  {Said}}, \bibinfo {author} {\bibfnamefont {T.}~\bibnamefont {Gog}}, \bibinfo
  {author} {\bibfnamefont {T.~F.}\ \bibnamefont {Qi}}, \bibinfo {author}
  {\bibfnamefont {G.}~\bibnamefont {Cao}}, \bibinfo {author} {\bibfnamefont
  {A.~M.}\ \bibnamefont {Tsvelik}}, \bibinfo {author} {\bibfnamefont
  {J.}~\bibnamefont {van~den Brink}},\ and\ \bibinfo {author} {\bibfnamefont
  {J.~P.}\ \bibnamefont {Hill}},\ }\bibfield  {title} {\bibinfo {title}
  {{Testing the Validity of the Strong Spin-Orbit-Coupling Limit for
  Octahedrally Coordinated Iridate Compounds in a Model System
  ${\mathrm{Sr}}_{3}{\mathrm{CuIrO}}_{6}$}},\ }\href
  {https://doi.org/10.1103/PhysRevLett.109.157401} {\bibfield  {journal}
  {\bibinfo  {journal} {Phys. Rev. Lett.}\ }\textbf {\bibinfo {volume} {109}},\
  \bibinfo {pages} {157401} (\bibinfo {year} {2012})}\BibitemShut {NoStop}%
\bibitem [{\citenamefont {Gretarsson}\ \emph {et~al.}(2013)\citenamefont
  {Gretarsson}, \citenamefont {Clancy}, \citenamefont {Liu}, \citenamefont
  {Hill}, \citenamefont {Bozin}, \citenamefont {Singh}, \citenamefont {Manni},
  \citenamefont {Gegenwart}, \citenamefont {Kim}, \citenamefont {Said},
  \citenamefont {Casa}, \citenamefont {Gog}, \citenamefont {Upton},
  \citenamefont {Kim}, \citenamefont {Yu}, \citenamefont {Katukuri},
  \citenamefont {Hozoi}, \citenamefont {van~den Brink},\ and\ \citenamefont
  {Kim}}]{Gretarsson2013PRL}%
  \BibitemOpen
  \bibfield  {author} {\bibinfo {author} {\bibfnamefont {H.}~\bibnamefont
  {Gretarsson}}, \bibinfo {author} {\bibfnamefont {J.~P.}\ \bibnamefont
  {Clancy}}, \bibinfo {author} {\bibfnamefont {X.}~\bibnamefont {Liu}},
  \bibinfo {author} {\bibfnamefont {J.~P.}\ \bibnamefont {Hill}}, \bibinfo
  {author} {\bibfnamefont {E.}~\bibnamefont {Bozin}}, \bibinfo {author}
  {\bibfnamefont {Y.}~\bibnamefont {Singh}}, \bibinfo {author} {\bibfnamefont
  {S.}~\bibnamefont {Manni}}, \bibinfo {author} {\bibfnamefont
  {P.}~\bibnamefont {Gegenwart}}, \bibinfo {author} {\bibfnamefont
  {J.}~\bibnamefont {Kim}}, \bibinfo {author} {\bibfnamefont {A.~H.}\
  \bibnamefont {Said}}, \bibinfo {author} {\bibfnamefont {D.}~\bibnamefont
  {Casa}}, \bibinfo {author} {\bibfnamefont {T.}~\bibnamefont {Gog}}, \bibinfo
  {author} {\bibfnamefont {M.~H.}\ \bibnamefont {Upton}}, \bibinfo {author}
  {\bibfnamefont {H.-S.}\ \bibnamefont {Kim}}, \bibinfo {author} {\bibfnamefont
  {J.}~\bibnamefont {Yu}}, \bibinfo {author} {\bibfnamefont {V.~M.}\
  \bibnamefont {Katukuri}}, \bibinfo {author} {\bibfnamefont {L.}~\bibnamefont
  {Hozoi}}, \bibinfo {author} {\bibfnamefont {J.}~\bibnamefont {van~den
  Brink}},\ and\ \bibinfo {author} {\bibfnamefont {Y.-J.}\ \bibnamefont
  {Kim}},\ }\bibfield  {title} {\bibinfo {title} {{Crystal-Field Splitting and
  Correlation Effect on the Electronic Structure of
  ${A}_{2}{\mathrm{IrO}}_{3}$}},\ }\href
  {https://doi.org/10.1103/PhysRevLett.110.076402} {\bibfield  {journal}
  {\bibinfo  {journal} {Phys. Rev. Lett.}\ }\textbf {\bibinfo {volume} {110}},\
  \bibinfo {pages} {076402} (\bibinfo {year} {2013})}\BibitemShut {NoStop}%
\bibitem [{\citenamefont {Dey}\ \emph {et~al.}(2014)\citenamefont {Dey},
  \citenamefont {Kumar}, \citenamefont {Mahajan}, \citenamefont {Kaushik},\
  and\ \citenamefont {Siruguri}}]{Dey2014PRB}%
  \BibitemOpen
  \bibfield  {author} {\bibinfo {author} {\bibfnamefont {T.}~\bibnamefont
  {Dey}}, \bibinfo {author} {\bibfnamefont {R.}~\bibnamefont {Kumar}}, \bibinfo
  {author} {\bibfnamefont {A.~V.}\ \bibnamefont {Mahajan}}, \bibinfo {author}
  {\bibfnamefont {S.~D.}\ \bibnamefont {Kaushik}},\ and\ \bibinfo {author}
  {\bibfnamefont {V.}~\bibnamefont {Siruguri}},\ }\bibfield  {title} {\bibinfo
  {title} {{Unconventional magnetism in the spin-orbit-driven Mott insulators
  ${\mathrm{Ba}}_{3}M$${\mathrm{Ir}}_{2}$${O}_{9}$ $(M=\mathrm{Sc},Y)$}},\
  }\href {https://doi.org/10.1103/PhysRevB.89.205101} {\bibfield  {journal}
  {\bibinfo  {journal} {Phys. Rev. B}\ }\textbf {\bibinfo {volume} {89}},\
  \bibinfo {pages} {205101} (\bibinfo {year} {2014})}\BibitemShut {NoStop}%
\bibitem [{\citenamefont {Terzic}\ \emph {et~al.}(2015)\citenamefont {Terzic},
  \citenamefont {Wang}, \citenamefont {Ye}, \citenamefont {Song}, \citenamefont
  {Yuan}, \citenamefont {Aswartham}, \citenamefont {DeLong}, \citenamefont
  {Streltsov}, \citenamefont {Khomskii},\ and\ \citenamefont
  {Cao}}]{Terzic2015}%
  \BibitemOpen
  \bibfield  {author} {\bibinfo {author} {\bibfnamefont {J.}~\bibnamefont
  {Terzic}}, \bibinfo {author} {\bibfnamefont {J.~C.}\ \bibnamefont {Wang}},
  \bibinfo {author} {\bibfnamefont {F.}~\bibnamefont {Ye}}, \bibinfo {author}
  {\bibfnamefont {W.~H.}\ \bibnamefont {Song}}, \bibinfo {author}
  {\bibfnamefont {S.~J.}\ \bibnamefont {Yuan}}, \bibinfo {author}
  {\bibfnamefont {S.}~\bibnamefont {Aswartham}}, \bibinfo {author}
  {\bibfnamefont {L.~E.}\ \bibnamefont {DeLong}}, \bibinfo {author}
  {\bibfnamefont {S.~V.}\ \bibnamefont {Streltsov}}, \bibinfo {author}
  {\bibfnamefont {D.~I.}\ \bibnamefont {Khomskii}},\ and\ \bibinfo {author}
  {\bibfnamefont {G.}~\bibnamefont {Cao}},\ }\bibfield  {title} {\bibinfo
  {title} {Coexisting charge and magnetic orders in the dimer-chain iridate
  $\mathrm{B}{\mathrm{a}}_{5}\mathrm{AlI}{\mathrm{r}}_{2}{\mathrm{o}}_{11}$},\
  }\href {https://doi.org/10.1103/PhysRevB.91.235147} {\bibfield  {journal}
  {\bibinfo  {journal} {Phys. Rev. B}\ }\textbf {\bibinfo {volume} {91}},\
  \bibinfo {pages} {235147} (\bibinfo {year} {2015})}\BibitemShut {NoStop}%
\bibitem [{\citenamefont {Hermann}\ \emph {et~al.}(2018)\citenamefont
  {Hermann}, \citenamefont {Altmeyer}, \citenamefont {Ebad-Allah},
  \citenamefont {Freund}, \citenamefont {Jesche}, \citenamefont {Tsirlin},
  \citenamefont {Hanfland}, \citenamefont {Gegenwart}, \citenamefont {Mazin},
  \citenamefont {Khomskii}, \citenamefont {Valent\'{\i}},\ and\ \citenamefont
  {Kuntscher}}]{Hermann2018}%
  \BibitemOpen
  \bibfield  {author} {\bibinfo {author} {\bibfnamefont {V.}~\bibnamefont
  {Hermann}}, \bibinfo {author} {\bibfnamefont {M.}~\bibnamefont {Altmeyer}},
  \bibinfo {author} {\bibfnamefont {J.}~\bibnamefont {Ebad-Allah}}, \bibinfo
  {author} {\bibfnamefont {F.}~\bibnamefont {Freund}}, \bibinfo {author}
  {\bibfnamefont {A.}~\bibnamefont {Jesche}}, \bibinfo {author} {\bibfnamefont
  {A.~A.}\ \bibnamefont {Tsirlin}}, \bibinfo {author} {\bibfnamefont
  {M.}~\bibnamefont {Hanfland}}, \bibinfo {author} {\bibfnamefont
  {P.}~\bibnamefont {Gegenwart}}, \bibinfo {author} {\bibfnamefont {I.~I.}\
  \bibnamefont {Mazin}}, \bibinfo {author} {\bibfnamefont {D.~I.}\ \bibnamefont
  {Khomskii}}, \bibinfo {author} {\bibfnamefont {R.}~\bibnamefont
  {Valent\'{\i}}},\ and\ \bibinfo {author} {\bibfnamefont {C.~A.}\ \bibnamefont
  {Kuntscher}},\ }\bibfield  {title} {\bibinfo {title} {{Competition between
  spin-orbit coupling, magnetism, and dimerization in the honeycomb iridates:
  $\ensuremath{\alpha}\ensuremath{-}{\mathrm{Li}}_{2}{\mathrm{IrO}}_{3}$ under
  pressure}},\ }\href {https://doi.org/10.1103/PhysRevB.97.020104} {\bibfield
  {journal} {\bibinfo  {journal} {Phys. Rev. B}\ }\textbf {\bibinfo {volume}
  {97}},\ \bibinfo {pages} {020104} (\bibinfo {year} {2018})}\BibitemShut
  {NoStop}%
\bibitem [{\citenamefont {Ye}\ \emph {et~al.}(2018)\citenamefont {Ye},
  \citenamefont {Kim}, \citenamefont {Kim}, \citenamefont {Won}, \citenamefont
  {Haule}, \citenamefont {Vanderbilt}, \citenamefont {Cheong},\ and\
  \citenamefont {Blumberg}}]{Ye2018PRB}%
  \BibitemOpen
  \bibfield  {author} {\bibinfo {author} {\bibfnamefont {M.}~\bibnamefont
  {Ye}}, \bibinfo {author} {\bibfnamefont {H.-S.}\ \bibnamefont {Kim}},
  \bibinfo {author} {\bibfnamefont {J.-W.}\ \bibnamefont {Kim}}, \bibinfo
  {author} {\bibfnamefont {C.-J.}\ \bibnamefont {Won}}, \bibinfo {author}
  {\bibfnamefont {K.}~\bibnamefont {Haule}}, \bibinfo {author} {\bibfnamefont
  {D.}~\bibnamefont {Vanderbilt}}, \bibinfo {author} {\bibfnamefont {S.-W.}\
  \bibnamefont {Cheong}},\ and\ \bibinfo {author} {\bibfnamefont
  {G.}~\bibnamefont {Blumberg}},\ }\bibfield  {title} {\bibinfo {title}
  {{Covalency-driven collapse of strong spin-orbit coupling in face-sharing
  iridium octahedra}},\ }\href {https://doi.org/10.1103/PhysRevB.98.201105}
  {\bibfield  {journal} {\bibinfo  {journal} {Phys. Rev. B}\ }\textbf {\bibinfo
  {volume} {98}},\ \bibinfo {pages} {201105} (\bibinfo {year}
  {2018})}\BibitemShut {NoStop}%
\bibitem [{\citenamefont {Wang}\ \emph {et~al.}(2019)\citenamefont {Wang},
  \citenamefont {Wang}, \citenamefont {Kim}, \citenamefont {Upton},
  \citenamefont {Casa}, \citenamefont {Gog}, \citenamefont {Cao}, \citenamefont
  {Kotliar}, \citenamefont {Dean},\ and\ \citenamefont {Liu}}]{Wang2019}%
  \BibitemOpen
  \bibfield  {author} {\bibinfo {author} {\bibfnamefont {Y.}~\bibnamefont
  {Wang}}, \bibinfo {author} {\bibfnamefont {R.}~\bibnamefont {Wang}}, \bibinfo
  {author} {\bibfnamefont {J.}~\bibnamefont {Kim}}, \bibinfo {author}
  {\bibfnamefont {M.~H.}\ \bibnamefont {Upton}}, \bibinfo {author}
  {\bibfnamefont {D.}~\bibnamefont {Casa}}, \bibinfo {author} {\bibfnamefont
  {T.}~\bibnamefont {Gog}}, \bibinfo {author} {\bibfnamefont {G.}~\bibnamefont
  {Cao}}, \bibinfo {author} {\bibfnamefont {G.}~\bibnamefont {Kotliar}},
  \bibinfo {author} {\bibfnamefont {M.~P.~M.}\ \bibnamefont {Dean}},\ and\
  \bibinfo {author} {\bibfnamefont {X.}~\bibnamefont {Liu}},\ }\bibfield
  {title} {\bibinfo {title} {{Direct Detection of Dimer Orbitals in
  ${\mathrm{Ba}}_{5}{\mathrm{AlIr}}_{2}{\mathrm{O}}_{11}$}},\ }\href
  {https://doi.org/10.1103/PhysRevLett.122.106401} {\bibfield  {journal}
  {\bibinfo  {journal} {Phys. Rev. Lett.}\ }\textbf {\bibinfo {volume} {122}},\
  \bibinfo {pages} {106401} (\bibinfo {year} {2019})}\BibitemShut {NoStop}%
\bibitem [{\citenamefont {Zhao}\ \emph {et~al.}(2019)\citenamefont {Zhao},
  \citenamefont {Ye}, \citenamefont {Zheng}, \citenamefont {Hu}, \citenamefont
  {Ni}, \citenamefont {Zhang}, \citenamefont {Kimchi},\ and\ \citenamefont
  {Cao}}]{Zhao2019PRB}%
  \BibitemOpen
  \bibfield  {author} {\bibinfo {author} {\bibfnamefont {H.}~\bibnamefont
  {Zhao}}, \bibinfo {author} {\bibfnamefont {F.}~\bibnamefont {Ye}}, \bibinfo
  {author} {\bibfnamefont {H.}~\bibnamefont {Zheng}}, \bibinfo {author}
  {\bibfnamefont {B.}~\bibnamefont {Hu}}, \bibinfo {author} {\bibfnamefont
  {Y.}~\bibnamefont {Ni}}, \bibinfo {author} {\bibfnamefont {Y.}~\bibnamefont
  {Zhang}}, \bibinfo {author} {\bibfnamefont {I.}~\bibnamefont {Kimchi}},\ and\
  \bibinfo {author} {\bibfnamefont {G.}~\bibnamefont {Cao}},\ }\bibfield
  {title} {\bibinfo {title} {{Ground state in the novel dimer iridate
  $\mathrm{B}{\mathrm{a}}_{13}\mathrm{I}{\mathrm{r}}_{6}{\mathrm{O}}_{30}$ with
  $\mathrm{I}{\mathrm{r}}^{6+}(5{d}^{3})$ ions}},\ }\href
  {https://doi.org/10.1103/PhysRevB.100.064418} {\bibfield  {journal} {\bibinfo
   {journal} {Phys. Rev. B}\ }\textbf {\bibinfo {volume} {100}},\ \bibinfo
  {pages} {064418} (\bibinfo {year} {2019})}\BibitemShut {NoStop}%
\bibitem [{\citenamefont {Khan}\ \emph {et~al.}(2019)\citenamefont {Khan},
  \citenamefont {Bandyopadhyay}, \citenamefont {Nag}, \citenamefont {Kumar},
  \citenamefont {Mahajan},\ and\ \citenamefont {Ray}}]{Khan2019PRB}%
  \BibitemOpen
  \bibfield  {author} {\bibinfo {author} {\bibfnamefont {M.~S.}\ \bibnamefont
  {Khan}}, \bibinfo {author} {\bibfnamefont {A.}~\bibnamefont {Bandyopadhyay}},
  \bibinfo {author} {\bibfnamefont {A.}~\bibnamefont {Nag}}, \bibinfo {author}
  {\bibfnamefont {V.}~\bibnamefont {Kumar}}, \bibinfo {author} {\bibfnamefont
  {A.~V.}\ \bibnamefont {Mahajan}},\ and\ \bibinfo {author} {\bibfnamefont
  {S.}~\bibnamefont {Ray}},\ }\bibfield  {title} {\bibinfo {title} {{Magnetic
  ground state of the distorted $6H$ perovskite
  ${\mathrm{Ba}}_{3}{\mathrm{CdIr}}_{2}{\mathrm{O}}_{9}$}},\ }\href
  {https://doi.org/10.1103/PhysRevB.100.064423} {\bibfield  {journal} {\bibinfo
   {journal} {Phys. Rev. B}\ }\textbf {\bibinfo {volume} {100}},\ \bibinfo
  {pages} {064423} (\bibinfo {year} {2019})}\BibitemShut {NoStop}%
\bibitem [{\citenamefont {Jeong}\ \emph {et~al.}(2020)\citenamefont {Jeong},
  \citenamefont {Lenz}, \citenamefont {Gukasov}, \citenamefont {Fabr\`eges},
  \citenamefont {Sazonov}, \citenamefont {Hutanu}, \citenamefont {Louat},
  \citenamefont {Bounoua}, \citenamefont {Martins}, \citenamefont {Biermann},
  \citenamefont {Brouet}, \citenamefont {Sidis},\ and\ \citenamefont
  {Bourges}}]{Jeong2020PRL}%
  \BibitemOpen
  \bibfield  {author} {\bibinfo {author} {\bibfnamefont {J.}~\bibnamefont
  {Jeong}}, \bibinfo {author} {\bibfnamefont {B.}~\bibnamefont {Lenz}},
  \bibinfo {author} {\bibfnamefont {A.}~\bibnamefont {Gukasov}}, \bibinfo
  {author} {\bibfnamefont {X.}~\bibnamefont {Fabr\`eges}}, \bibinfo {author}
  {\bibfnamefont {A.}~\bibnamefont {Sazonov}}, \bibinfo {author} {\bibfnamefont
  {V.}~\bibnamefont {Hutanu}}, \bibinfo {author} {\bibfnamefont
  {A.}~\bibnamefont {Louat}}, \bibinfo {author} {\bibfnamefont
  {D.}~\bibnamefont {Bounoua}}, \bibinfo {author} {\bibfnamefont
  {C.}~\bibnamefont {Martins}}, \bibinfo {author} {\bibfnamefont
  {S.}~\bibnamefont {Biermann}}, \bibinfo {author} {\bibfnamefont
  {V.}~\bibnamefont {Brouet}}, \bibinfo {author} {\bibfnamefont
  {Y.}~\bibnamefont {Sidis}},\ and\ \bibinfo {author} {\bibfnamefont
  {P.}~\bibnamefont {Bourges}},\ }\bibfield  {title} {\bibinfo {title}
  {{Magnetization Density Distribution of
  ${\mathrm{Sr}}_{2}{\mathrm{IrO}}_{4}$: Deviation from a Local
  ${J}_{\mathrm{eff}}=1/2$ Picture}},\ }\href
  {https://doi.org/10.1103/PhysRevLett.125.097202} {\bibfield  {journal}
  {\bibinfo  {journal} {Phys. Rev. Lett.}\ }\textbf {\bibinfo {volume} {125}},\
  \bibinfo {pages} {097202} (\bibinfo {year} {2020})}\BibitemShut {NoStop}%
\bibitem [{\citenamefont {Kumar}\ \emph {et~al.}(2021)\citenamefont {Kumar},
  \citenamefont {Panda}, \citenamefont {Patidar}, \citenamefont {Ojha},
  \citenamefont {Mandal}, \citenamefont {Das}, \citenamefont {Freeland},
  \citenamefont {Ganesan}, \citenamefont {Baker},\ and\ \citenamefont
  {Middey}}]{Kumar2021PRB}%
  \BibitemOpen
  \bibfield  {author} {\bibinfo {author} {\bibfnamefont {S.}~\bibnamefont
  {Kumar}}, \bibinfo {author} {\bibfnamefont {S.~K.}\ \bibnamefont {Panda}},
  \bibinfo {author} {\bibfnamefont {M.~M.}\ \bibnamefont {Patidar}}, \bibinfo
  {author} {\bibfnamefont {S.~K.}\ \bibnamefont {Ojha}}, \bibinfo {author}
  {\bibfnamefont {P.}~\bibnamefont {Mandal}}, \bibinfo {author} {\bibfnamefont
  {G.}~\bibnamefont {Das}}, \bibinfo {author} {\bibfnamefont {J.~W.}\
  \bibnamefont {Freeland}}, \bibinfo {author} {\bibfnamefont {V.}~\bibnamefont
  {Ganesan}}, \bibinfo {author} {\bibfnamefont {P.~J.}\ \bibnamefont {Baker}},\
  and\ \bibinfo {author} {\bibfnamefont {S.}~\bibnamefont {Middey}},\
  }\bibfield  {title} {\bibinfo {title} {{Spin-liquid behavior of the
  three-dimensional magnetic system
  ${\mathrm{Ba}}_{3}{\mathrm{NiIr}}_{2}{\mathrm{O}}_{9}$ with $S=1$}},\ }\href
  {https://doi.org/10.1103/PhysRevB.103.184405} {\bibfield  {journal} {\bibinfo
   {journal} {Phys. Rev. B}\ }\textbf {\bibinfo {volume} {103}},\ \bibinfo
  {pages} {184405} (\bibinfo {year} {2021})}\BibitemShut {NoStop}%
\bibitem [{\citenamefont {Lane}\ \emph {et~al.}(2020)\citenamefont {Lane},
  \citenamefont {Zhang}, \citenamefont {Furness}, \citenamefont {Markiewicz},
  \citenamefont {Barbiellini}, \citenamefont {Sun},\ and\ \citenamefont
  {Bansil}}]{Lane2020PRB}%
  \BibitemOpen
  \bibfield  {author} {\bibinfo {author} {\bibfnamefont {C.}~\bibnamefont
  {Lane}}, \bibinfo {author} {\bibfnamefont {Y.}~\bibnamefont {Zhang}},
  \bibinfo {author} {\bibfnamefont {J.~W.}\ \bibnamefont {Furness}}, \bibinfo
  {author} {\bibfnamefont {R.~S.}\ \bibnamefont {Markiewicz}}, \bibinfo
  {author} {\bibfnamefont {B.}~\bibnamefont {Barbiellini}}, \bibinfo {author}
  {\bibfnamefont {J.}~\bibnamefont {Sun}},\ and\ \bibinfo {author}
  {\bibfnamefont {A.}~\bibnamefont {Bansil}},\ }\bibfield  {title} {\bibinfo
  {title} {{First-principles calculation of spin and orbital contributions to
  magnetically ordered moments in ${\mathrm{Sr}}_{2}{\mathrm{IrO}}_{4}$}},\
  }\href {https://doi.org/10.1103/PhysRevB.101.155110} {\bibfield  {journal}
  {\bibinfo  {journal} {Phys. Rev. B}\ }\textbf {\bibinfo {volume} {101}},\
  \bibinfo {pages} {155110} (\bibinfo {year} {2020})}\BibitemShut {NoStop}%
\bibitem [{\citenamefont {Fujioka}\ \emph {et~al.}(2018)\citenamefont
  {Fujioka}, \citenamefont {Okawa}, \citenamefont {Masuko}, \citenamefont
  {Yamamoto},\ and\ \citenamefont {Tokura}}]{Fujioka2018}%
  \BibitemOpen
  \bibfield  {author} {\bibinfo {author} {\bibfnamefont {J.}~\bibnamefont
  {Fujioka}}, \bibinfo {author} {\bibfnamefont {T.}~\bibnamefont {Okawa}},
  \bibinfo {author} {\bibfnamefont {M.}~\bibnamefont {Masuko}}, \bibinfo
  {author} {\bibfnamefont {A.}~\bibnamefont {Yamamoto}},\ and\ \bibinfo
  {author} {\bibfnamefont {Y.}~\bibnamefont {Tokura}},\ }\bibfield  {title}
  {\bibinfo {title} {{Charge dynamics and metal-insulator transition in
  perovskite SrIr$_{1-x}$Sn$_x$O$_3$}},\ }\href
  {https://doi.org/10.7566/JPSJ.87.123706} {\bibfield  {journal} {\bibinfo
  {journal} {Journal of the Physical Society of Japan}\ }\textbf {\bibinfo
  {volume} {87}},\ \bibinfo {pages} {123706} (\bibinfo {year}
  {2018})}\BibitemShut {NoStop}%
\bibitem [{\citenamefont {Yang}\ \emph {et~al.}(2019)\citenamefont {Yang},
  \citenamefont {Wang}, \citenamefont {Zhen}, \citenamefont {Ma}, \citenamefont
  {Ling}, \citenamefont {Tong}, \citenamefont {Zhang}, \citenamefont {Pi},\
  and\ \citenamefont {Zhu}}]{Yang2019}%
  \BibitemOpen
  \bibfield  {author} {\bibinfo {author} {\bibfnamefont {J.}~\bibnamefont
  {Yang}}, \bibinfo {author} {\bibfnamefont {J.~R.}\ \bibnamefont {Wang}},
  \bibinfo {author} {\bibfnamefont {W.~L.}\ \bibnamefont {Zhen}}, \bibinfo
  {author} {\bibfnamefont {L.}~\bibnamefont {Ma}}, \bibinfo {author}
  {\bibfnamefont {L.~S.}\ \bibnamefont {Ling}}, \bibinfo {author}
  {\bibfnamefont {W.}~\bibnamefont {Tong}}, \bibinfo {author} {\bibfnamefont
  {C.~J.}\ \bibnamefont {Zhang}}, \bibinfo {author} {\bibfnamefont
  {L.}~\bibnamefont {Pi}},\ and\ \bibinfo {author} {\bibfnamefont {W.~K.}\
  \bibnamefont {Zhu}},\ }\bibfield  {title} {\bibinfo {title}
  {{Frustration-induced non-Curie-Weiss paramagnetism in
  ${\mathrm{La}}_{3}{\mathrm{Ir}}_{3}{\mathrm{O}}_{11}$: A fractional valence
  state iridate}},\ }\href {https://doi.org/10.1103/PhysRevB.100.205107}
  {\bibfield  {journal} {\bibinfo  {journal} {Phys. Rev. B}\ }\textbf {\bibinfo
  {volume} {100}},\ \bibinfo {pages} {205107} (\bibinfo {year}
  {2019})}\BibitemShut {NoStop}%
\bibitem [{\citenamefont {Aoyama}\ \emph {et~al.}(2019)\citenamefont {Aoyama},
  \citenamefont {Emi}, \citenamefont {Tabata}, \citenamefont {Nambu},
  \citenamefont {Nakao}, \citenamefont {Yamauchi},\ and\ \citenamefont
  {Ohgushi}}]{Aoyama2019}%
  \BibitemOpen
  \bibfield  {author} {\bibinfo {author} {\bibfnamefont {T.}~\bibnamefont
  {Aoyama}}, \bibinfo {author} {\bibfnamefont {K.}~\bibnamefont {Emi}},
  \bibinfo {author} {\bibfnamefont {C.}~\bibnamefont {Tabata}}, \bibinfo
  {author} {\bibfnamefont {Y.}~\bibnamefont {Nambu}}, \bibinfo {author}
  {\bibfnamefont {H.}~\bibnamefont {Nakao}}, \bibinfo {author} {\bibfnamefont
  {T.}~\bibnamefont {Yamauchi}},\ and\ \bibinfo {author} {\bibfnamefont
  {K.}~\bibnamefont {Ohgushi}},\ }\bibfield  {title} {\bibinfo {title} {{A
  Semimetallic State in La$_3$Ir$_3$O$_{11}$ with the KSbO$_3$ Structure}},\
  }\href {https://doi.org/10.7566/JPSJ.88.093706} {\bibfield  {journal}
  {\bibinfo  {journal} {Journal of the Physical Society of Japan}\ }\textbf
  {\bibinfo {volume} {88}},\ \bibinfo {pages} {093706} (\bibinfo {year}
  {2019})}\BibitemShut {NoStop}%
\bibitem [{\citenamefont {Abraham}\ \emph {et~al.}(1982)\citenamefont
  {Abraham}, \citenamefont {Trehoux}, \citenamefont {Thomas},\ and\
  \citenamefont {Wagner}}]{Abraham1982}%
  \BibitemOpen
  \bibfield  {author} {\bibinfo {author} {\bibfnamefont {F.}~\bibnamefont
  {Abraham}}, \bibinfo {author} {\bibfnamefont {J.}~\bibnamefont {Trehoux}},
  \bibinfo {author} {\bibfnamefont {D.}~\bibnamefont {Thomas}},\ and\ \bibinfo
  {author} {\bibfnamefont {F.}~\bibnamefont {Wagner}},\ }\bibfield  {title}
  {\bibinfo {title} {{Propri\'{e}t\'{e}s \'{e}lectriques, magn\'{e}tiques et de
  r\'{e}sonance M\"{o}ssbauer de La$_3$Ir$_3$O$_{11}$}},\ }\href
  {https://doi.org/https://doi.org/10.1016/0022-5088(82)90149-7} {\bibfield
  {journal} {\bibinfo  {journal} {Journal of the Less Common Metals}\ }\textbf
  {\bibinfo {volume} {84}},\ \bibinfo {pages} {245} (\bibinfo {year}
  {1982})}\BibitemShut {NoStop}%
\bibitem [{Note1()}]{Note1}%
  \BibitemOpen
  \bibinfo {note} {See Supplemental Materials at \protect \url
  {http://link.aps.org/supplemental/xxx} for the details of sample synthesis,
  experimental methods, and details of Drude-Lorentz analysis and theoretical
  calculations,which includes Refs.~\cite
  {Homes1993,Dressel2002,Kresse1996PRB,Kresse1996,Kresse1999,Blochl1994,Perdew1996,Dudarev1998,Marzari2012,Marrazzo2024}}\BibitemShut
  {NoStop}%
\bibitem [{\citenamefont {Homes}\ \emph {et~al.}(1993)\citenamefont {Homes},
  \citenamefont {Reedyk}, \citenamefont {Cradles},\ and\ \citenamefont
  {Timusk}}]{Homes1993}%
  \BibitemOpen
  \bibfield  {author} {\bibinfo {author} {\bibfnamefont {C.~C.}\ \bibnamefont
  {Homes}}, \bibinfo {author} {\bibfnamefont {M.}~\bibnamefont {Reedyk}},
  \bibinfo {author} {\bibfnamefont {D.~A.}\ \bibnamefont {Cradles}},\ and\
  \bibinfo {author} {\bibfnamefont {T.}~\bibnamefont {Timusk}},\ }\bibfield
  {title} {\bibinfo {title} {{Technique for measuring the reflectance of
  irregular, submillimeter-sized samples}},\ }\href
  {https://doi.org/10.1364/AO.32.002976} {\bibfield  {journal} {\bibinfo
  {journal} {Appl. Opt.}\ }\textbf {\bibinfo {volume} {32}},\ \bibinfo {pages}
  {2976} (\bibinfo {year} {1993})}\BibitemShut {NoStop}%
\bibitem [{\citenamefont {Dressel}\ and\ \citenamefont
  {Gr\"uner}(2002)}]{Dressel2002}%
  \BibitemOpen
  \bibfield  {author} {\bibinfo {author} {\bibfnamefont {M.}~\bibnamefont
  {Dressel}}\ and\ \bibinfo {author} {\bibfnamefont {G.}~\bibnamefont
  {Gr\"uner}},\ }\href@noop {} {\emph {\bibinfo {title} {{Electrodynamics of
  Solids}}}}\ (\bibinfo  {publisher} {Cambridge University press},\ \bibinfo
  {year} {2002})\BibitemShut {NoStop}%
\bibitem [{\citenamefont {Kresse}\ and\ \citenamefont
  {Furthm\"uller}(1996)}]{Kresse1996PRB}%
  \BibitemOpen
  \bibfield  {author} {\bibinfo {author} {\bibfnamefont {G.}~\bibnamefont
  {Kresse}}\ and\ \bibinfo {author} {\bibfnamefont {J.}~\bibnamefont
  {Furthm\"uller}},\ }\bibfield  {title} {\bibinfo {title} {{Efficient
  iterative schemes for ab initio total-energy calculations using a plane-wave
  basis set}},\ }\href {https://doi.org/10.1103/PhysRevB.54.11169} {\bibfield
  {journal} {\bibinfo  {journal} {Phys. Rev. B}\ }\textbf {\bibinfo {volume}
  {54}},\ \bibinfo {pages} {11169} (\bibinfo {year} {1996})}\BibitemShut
  {NoStop}%
\bibitem [{\citenamefont {Kresse}\ and\ \citenamefont
  {Furthm\"{u}ller}(1996)}]{Kresse1996}%
  \BibitemOpen
  \bibfield  {author} {\bibinfo {author} {\bibfnamefont {G.}~\bibnamefont
  {Kresse}}\ and\ \bibinfo {author} {\bibfnamefont {J.}~\bibnamefont
  {Furthm\"{u}ller}},\ }\bibfield  {title} {\bibinfo {title} {{Efficiency of
  ab-initio total energy calculations for metals and semiconductors using a
  plane-wave basis set}},\ }\href
  {https://doi.org/https://doi.org/10.1016/0927-0256(96)00008-0} {\bibfield
  {journal} {\bibinfo  {journal} {Computational Materials Science}\ }\textbf
  {\bibinfo {volume} {6}},\ \bibinfo {pages} {15} (\bibinfo {year}
  {1996})}\BibitemShut {NoStop}%
\bibitem [{\citenamefont {Kresse}\ and\ \citenamefont
  {Joubert}(1999)}]{Kresse1999}%
  \BibitemOpen
  \bibfield  {author} {\bibinfo {author} {\bibfnamefont {G.}~\bibnamefont
  {Kresse}}\ and\ \bibinfo {author} {\bibfnamefont {D.}~\bibnamefont
  {Joubert}},\ }\bibfield  {title} {\bibinfo {title} {{From ultrasoft
  pseudopotentials to the projector augmented-wave method}},\ }\href
  {https://doi.org/10.1103/PhysRevB.59.1758} {\bibfield  {journal} {\bibinfo
  {journal} {Phys. Rev. B}\ }\textbf {\bibinfo {volume} {59}},\ \bibinfo
  {pages} {1758} (\bibinfo {year} {1999})}\BibitemShut {NoStop}%
\bibitem [{\citenamefont {Bl\"ochl}(1994)}]{Blochl1994}%
  \BibitemOpen
  \bibfield  {author} {\bibinfo {author} {\bibfnamefont {P.~E.}\ \bibnamefont
  {Bl\"ochl}},\ }\bibfield  {title} {\bibinfo {title} {{Projector
  augmented-wave method}},\ }\href {https://doi.org/10.1103/PhysRevB.50.17953}
  {\bibfield  {journal} {\bibinfo  {journal} {Phys. Rev. B}\ }\textbf {\bibinfo
  {volume} {50}},\ \bibinfo {pages} {17953} (\bibinfo {year}
  {1994})}\BibitemShut {NoStop}%
\bibitem [{\citenamefont {Perdew}\ \emph {et~al.}(1996)\citenamefont {Perdew},
  \citenamefont {Burke},\ and\ \citenamefont {Ernzerhof}}]{Perdew1996}%
  \BibitemOpen
  \bibfield  {author} {\bibinfo {author} {\bibfnamefont {J.~P.}\ \bibnamefont
  {Perdew}}, \bibinfo {author} {\bibfnamefont {K.}~\bibnamefont {Burke}},\ and\
  \bibinfo {author} {\bibfnamefont {M.}~\bibnamefont {Ernzerhof}},\ }\bibfield
  {title} {\bibinfo {title} {{Generalized Gradient Approximation Made
  Simple}},\ }\href {https://doi.org/10.1103/PhysRevLett.77.3865} {\bibfield
  {journal} {\bibinfo  {journal} {Phys. Rev. Lett.}\ }\textbf {\bibinfo
  {volume} {77}},\ \bibinfo {pages} {3865} (\bibinfo {year}
  {1996})}\BibitemShut {NoStop}%
\bibitem [{\citenamefont {Dudarev}\ \emph {et~al.}(1998)\citenamefont
  {Dudarev}, \citenamefont {Botton}, \citenamefont {Savrasov}, \citenamefont
  {Humphreys},\ and\ \citenamefont {Sutton}}]{Dudarev1998}%
  \BibitemOpen
  \bibfield  {author} {\bibinfo {author} {\bibfnamefont {S.~L.}\ \bibnamefont
  {Dudarev}}, \bibinfo {author} {\bibfnamefont {G.~A.}\ \bibnamefont {Botton}},
  \bibinfo {author} {\bibfnamefont {S.~Y.}\ \bibnamefont {Savrasov}}, \bibinfo
  {author} {\bibfnamefont {C.~J.}\ \bibnamefont {Humphreys}},\ and\ \bibinfo
  {author} {\bibfnamefont {A.~P.}\ \bibnamefont {Sutton}},\ }\bibfield  {title}
  {\bibinfo {title} {{Electron-energy-loss spectra and the structural stability
  of nickel oxide: An LSDA+U study}},\ }\href
  {https://doi.org/10.1103/PhysRevB.57.1505} {\bibfield  {journal} {\bibinfo
  {journal} {Phys. Rev. B}\ }\textbf {\bibinfo {volume} {57}},\ \bibinfo
  {pages} {1505} (\bibinfo {year} {1998})}\BibitemShut {NoStop}%
\bibitem [{\citenamefont {Marzari}\ \emph {et~al.}(2012)\citenamefont
  {Marzari}, \citenamefont {Mostofi}, \citenamefont {Yates}, \citenamefont
  {Souza},\ and\ \citenamefont {Vanderbilt}}]{Marzari2012}%
  \BibitemOpen
  \bibfield  {author} {\bibinfo {author} {\bibfnamefont {N.}~\bibnamefont
  {Marzari}}, \bibinfo {author} {\bibfnamefont {A.~A.}\ \bibnamefont
  {Mostofi}}, \bibinfo {author} {\bibfnamefont {J.~R.}\ \bibnamefont {Yates}},
  \bibinfo {author} {\bibfnamefont {I.}~\bibnamefont {Souza}},\ and\ \bibinfo
  {author} {\bibfnamefont {D.}~\bibnamefont {Vanderbilt}},\ }\bibfield  {title}
  {\bibinfo {title} {{Maximally localized Wannier functions: Theory and
  applications}},\ }\href {https://doi.org/10.1103/RevModPhys.84.1419}
  {\bibfield  {journal} {\bibinfo  {journal} {Rev. Mod. Phys.}\ }\textbf
  {\bibinfo {volume} {84}},\ \bibinfo {pages} {1419} (\bibinfo {year}
  {2012})}\BibitemShut {NoStop}%
\bibitem [{\citenamefont {Marrazzo}\ \emph {et~al.}(2024)\citenamefont
  {Marrazzo}, \citenamefont {Beck}, \citenamefont {Margine}, \citenamefont
  {Marzari}, \citenamefont {Mostofi}, \citenamefont {Qiao}, \citenamefont
  {Souza}, \citenamefont {Tsirkin}, \citenamefont {Yates},\ and\ \citenamefont
  {Pizzi}}]{Marrazzo2024}%
  \BibitemOpen
  \bibfield  {author} {\bibinfo {author} {\bibfnamefont {A.}~\bibnamefont
  {Marrazzo}}, \bibinfo {author} {\bibfnamefont {S.}~\bibnamefont {Beck}},
  \bibinfo {author} {\bibfnamefont {E.~R.}\ \bibnamefont {Margine}}, \bibinfo
  {author} {\bibfnamefont {N.}~\bibnamefont {Marzari}}, \bibinfo {author}
  {\bibfnamefont {A.~A.}\ \bibnamefont {Mostofi}}, \bibinfo {author}
  {\bibfnamefont {J.}~\bibnamefont {Qiao}}, \bibinfo {author} {\bibfnamefont
  {I.}~\bibnamefont {Souza}}, \bibinfo {author} {\bibfnamefont {S.~S.}\
  \bibnamefont {Tsirkin}}, \bibinfo {author} {\bibfnamefont {J.~R.}\
  \bibnamefont {Yates}},\ and\ \bibinfo {author} {\bibfnamefont
  {G.}~\bibnamefont {Pizzi}},\ }\bibfield  {title} {\bibinfo {title}
  {{Wannier-function software ecosystem for materials simulations}},\ }\href
  {https://doi.org/10.1103/RevModPhys.96.045008} {\bibfield  {journal}
  {\bibinfo  {journal} {Rev. Mod. Phys.}\ }\textbf {\bibinfo {volume} {96}},\
  \bibinfo {pages} {045008} (\bibinfo {year} {2024})}\BibitemShut {NoStop}%
\bibitem [{\citenamefont {Basov}\ and\ \citenamefont
  {Timusk}(2005)}]{Basov2005}%
  \BibitemOpen
  \bibfield  {author} {\bibinfo {author} {\bibfnamefont {D.~N.}\ \bibnamefont
  {Basov}}\ and\ \bibinfo {author} {\bibfnamefont {T.}~\bibnamefont {Timusk}},\
  }\bibfield  {title} {\bibinfo {title} {{Electrodynamics of high-${T}_{c}$
  superconductors}},\ }\href {https://doi.org/10.1103/RevModPhys.77.721}
  {\bibfield  {journal} {\bibinfo  {journal} {Rev. Mod. Phys.}\ }\textbf
  {\bibinfo {volume} {77}},\ \bibinfo {pages} {721} (\bibinfo {year}
  {2005})}\BibitemShut {NoStop}%
\bibitem [{\citenamefont {Basov}\ \emph {et~al.}(2011)\citenamefont {Basov},
  \citenamefont {Averitt}, \citenamefont {van~der Marel}, \citenamefont
  {Dressel},\ and\ \citenamefont {Haule}}]{Basov2011}%
  \BibitemOpen
  \bibfield  {author} {\bibinfo {author} {\bibfnamefont {D.~N.}\ \bibnamefont
  {Basov}}, \bibinfo {author} {\bibfnamefont {R.~D.}\ \bibnamefont {Averitt}},
  \bibinfo {author} {\bibfnamefont {D.}~\bibnamefont {van~der Marel}}, \bibinfo
  {author} {\bibfnamefont {M.}~\bibnamefont {Dressel}},\ and\ \bibinfo {author}
  {\bibfnamefont {K.}~\bibnamefont {Haule}},\ }\bibfield  {title} {\bibinfo
  {title} {{Electrodynamics of correlated electron materials}},\ }\href
  {https://doi.org/10.1103/RevModPhys.83.471} {\bibfield  {journal} {\bibinfo
  {journal} {Rev. Mod. Phys.}\ }\textbf {\bibinfo {volume} {83}},\ \bibinfo
  {pages} {471} (\bibinfo {year} {2011})}\BibitemShut {NoStop}%
\bibitem [{\citenamefont {Zhang}\ \emph {et~al.}(2013)\citenamefont {Zhang},
  \citenamefont {Haule},\ and\ \citenamefont {Vanderbilt}}]{Zhang2013}%
  \BibitemOpen
  \bibfield  {author} {\bibinfo {author} {\bibfnamefont {H.}~\bibnamefont
  {Zhang}}, \bibinfo {author} {\bibfnamefont {K.}~\bibnamefont {Haule}},\ and\
  \bibinfo {author} {\bibfnamefont {D.}~\bibnamefont {Vanderbilt}},\ }\bibfield
   {title} {\bibinfo {title} {{Effective $J\mathbf{=}1/2$ Insulating State in
  Ruddlesden-Popper Iridates: An $\mathrm{LDA}\mathbf{+}\mathrm{DMFT}$
  Study}},\ }\href {https://doi.org/10.1103/PhysRevLett.111.246402} {\bibfield
  {journal} {\bibinfo  {journal} {Phys. Rev. Lett.}\ }\textbf {\bibinfo
  {volume} {111}},\ \bibinfo {pages} {246402} (\bibinfo {year}
  {2013})}\BibitemShut {NoStop}%
\bibitem [{\citenamefont {Phillips}(2010)}]{Phillips2010}%
  \BibitemOpen
  \bibfield  {author} {\bibinfo {author} {\bibfnamefont {P.}~\bibnamefont
  {Phillips}},\ }\bibfield  {title} {\bibinfo {title} {{Colloquium: Identifying
  the propagating charge modes in doped Mott insulators}},\ }\href
  {https://doi.org/10.1103/RevModPhys.82.1719} {\bibfield  {journal} {\bibinfo
  {journal} {Rev. Mod. Phys.}\ }\textbf {\bibinfo {volume} {82}},\ \bibinfo
  {pages} {1719} (\bibinfo {year} {2010})}\BibitemShut {NoStop}%
\bibitem [{\citenamefont {van~der Marel}\ and\ \citenamefont
  {Sawatzky}(1988)}]{Dirk1988}%
  \BibitemOpen
  \bibfield  {author} {\bibinfo {author} {\bibfnamefont {D.}~\bibnamefont
  {van~der Marel}}\ and\ \bibinfo {author} {\bibfnamefont {G.~A.}\ \bibnamefont
  {Sawatzky}},\ }\bibfield  {title} {\bibinfo {title} {Electron-electron
  interaction and localization in d and $f$ transition metals},\ }\href
  {https://doi.org/10.1103/PhysRevB.37.10674} {\bibfield  {journal} {\bibinfo
  {journal} {Phys. Rev. B}\ }\textbf {\bibinfo {volume} {37}},\ \bibinfo
  {pages} {10674} (\bibinfo {year} {1988})}\BibitemShut {NoStop}%
\bibitem [{\citenamefont {Ueda}\ \emph {et~al.}(2016)\citenamefont {Ueda},
  \citenamefont {Fujioka},\ and\ \citenamefont {Tokura}}]{Ueda2016PRB}%
  \BibitemOpen
  \bibfield  {author} {\bibinfo {author} {\bibfnamefont {K.}~\bibnamefont
  {Ueda}}, \bibinfo {author} {\bibfnamefont {J.}~\bibnamefont {Fujioka}},\ and\
  \bibinfo {author} {\bibfnamefont {Y.}~\bibnamefont {Tokura}},\ }\bibfield
  {title} {\bibinfo {title} {{Variation of optical conductivity spectra in the
  course of bandwidth-controlled metal-insulator transitions in pyrochlore
  iridates}},\ }\href {https://doi.org/10.1103/PhysRevB.93.245120} {\bibfield
  {journal} {\bibinfo  {journal} {Phys. Rev. B}\ }\textbf {\bibinfo {volume}
  {93}},\ \bibinfo {pages} {245120} (\bibinfo {year} {2016})}\BibitemShut
  {NoStop}%
\bibitem [{\citenamefont {Chen}\ \emph {et~al.}()\citenamefont {Chen},
  \citenamefont {Tian}, \citenamefont {He},\ and\ \citenamefont
  {Lu}}]{Chen2025}%
  \BibitemOpen
  \bibfield  {author} {\bibinfo {author} {\bibfnamefont {Y.}~\bibnamefont
  {Chen}}, \bibinfo {author} {\bibfnamefont {Y.-H.}\ \bibnamefont {Tian}},
  \bibinfo {author} {\bibfnamefont {R.-Q.}\ \bibnamefont {He}},\ and\ \bibinfo
  {author} {\bibfnamefont {Z.-Y.}\ \bibnamefont {Lu}},\ }\bibfield  {title}
  {\bibinfo {title} {{Spin-orbit coupling effects on orbital-selective
  correlations in a three-orbital model}},\ }\Eprint
  {https://arxiv.org/abs/arXiv:2503.14435v1 (2025)} {arXiv:2503.14435v1 (2025)}
  \BibitemShut {NoStop}%
\bibitem [{\citenamefont {Li}\ \emph {et~al.}(2025)\citenamefont {Li},
  \citenamefont {Guo}, \citenamefont {Xie}, \citenamefont {Wang},\ and\
  \citenamefont {Wang}}]{Li2025}%
  \BibitemOpen
  \bibfield  {author} {\bibinfo {author} {\bibfnamefont {R.-S.}\ \bibnamefont
  {Li}}, \bibinfo {author} {\bibfnamefont {R.}~\bibnamefont {Guo}}, \bibinfo
  {author} {\bibfnamefont {Z.}~\bibnamefont {Xie}}, \bibinfo {author}
  {\bibfnamefont {J.-T.}\ \bibnamefont {Wang}},\ and\ \bibinfo {author}
  {\bibfnamefont {F.}~\bibnamefont {Wang}},\ }\bibfield  {title} {\bibinfo
  {title} {{Spin-orbit-controlled correlation physics and orbital-selective
  electronic transitions in uranium monoxide revealed by many-body
  calculations}},\ }\href {https://doi.org/10.1039/d5cp01598g} {\bibfield
  {journal} {\bibinfo  {journal} {Phys. Chem. Chem. Phys.}\ }\textbf {\bibinfo
  {volume} {27}},\ \bibinfo {pages} {15508} (\bibinfo {year}
  {2025})}\BibitemShut {NoStop}%
\end{thebibliography}
\end{document}